\documentclass[12pt,usenatbib]{emulateapj}
\usepackage[hyperfootnotes=false,naturalnames=true,letterpaper,pdfstartview=FitH,pdfpagemode=none]{hyperref}
\shorttitle{Identifying Anomalies in Gravitational Lens Time Delays}
\shortauthors{Congdon, Keeton, and Nordgren}

\slugcomment{In press at The Astrophysical Journal}

\newcommand\Rfold{R_{\rm fold}}
\newcommand\Rcusp{R_{\rm cusp}}
\newcommand\Afold{A_{\rm fold}}
\newcommand\Acusp{A_{\rm cusp}}
\newcommand\Msun{M_{\odot}}
\newcommand\vect[1]{{\mbox{\boldmath $#1$}}}
\newcommand\btheta{\vect{\theta}}
\newcommand\bbeta{\vect{\beta}}

\begin{document}

\title{Identifying Anomalies in Gravitational Lens Time Delays}

\author{
  Arthur B. Congdon\altaffilmark{1,2},
  Charles R. Keeton\altaffilmark{2},
  and
  C. Erik Nordgren\altaffilmark{3}
}

\altaffiltext{1}{
Jet Propulsion Laboratory, California Institute of Technology,
4800 Oak Grove Drive, MS 169-327, Pasadena, CA 91109-8001;
acongdon@jpl.nasa.gov
}
\altaffiltext{2}{
Department of Physics and Astronomy, Rutgers University,
136 Frelinghuysen Road, Piscataway, NJ 08854-8019;
keeton@physics.rutgers.edu
}
\altaffiltext{3}{
Department of Chemistry, University of Pennsylvania,
231 South 34th Street, Philadelphia, PA 19104-6323;
nordgren@sas.upenn.edu
}

\begin{abstract}
We examine the ability of gravitational lens time delays to reveal complex structure in lens potentials.  In a previous paper, we predicted how the time delay between the bright pair of images in a ``fold'' lens scales with the image separation, for smooth lens potentials. Here we show that the proportionality constant increases with the quadrupole moment of the lens potential, and depends only weakly on the position of the source along the caustic. We use Monte Carlo simulations to determine the range of time delays that can be produced by realistic smooth lens models consisting of isothermal ellipsoid galaxies with tidal shear. We can then identify outliers as ``time delay anomalies.'' We find evidence for anomalies in close image pairs in the cusp lenses RX J1131$-$1231 and B1422+231. The anomalies in RX J1131$-$1231 provide strong evidence for substructure in the lens potential, while at this point the apparent anomalies in B1422+231 mainly indicate that the time delay measurements need to be improved.  We also find evidence for time delay anomalies in larger-separation image pairs in the fold lenses, B1608+656 and WFI 2033$-$4723, and the cusp lens RX J0911+0551. We suggest that these anomalies are caused by some combination of substructure and a complex lens environment. Finally, to assist future monitoring campaigns we use our smooth models with shear to predict the time delays for all known four-image lenses.
\end{abstract}


\section{INTRODUCTION}

Gravitational lensing has become a valuable probe of dark matter substructure in distant galaxies (see \S B.8 of \citealt{Kochanek_review}, and references therein). A growing body of evidence suggests that anomalous flux ratios, which are observed in many four-image quasar lenses \citep{MetcalfZhao,Keeton-cusp,Keeton-fold}, can be explained if a few percent of the projected mass within each lens galaxy's Einstein angle (which sets the spatial scale for lensing) is contained in cold dark matter (CDM) ``clumps'' \citep{Mao-flux,Metcalf-substruc,Chiba_substruc,Dalal-substruc,Kochanek-substruc}. It is then natural to ask whether it is possible to use lensing to measure properties beyond the mean substructure mass fraction.  Of particular interest is the clump mass function, because the discrepancy between the observed number of Local Group satellite galaxies and the theoretically predicted abundance of dark matter clumps varies strongly with mass \citep[e.g.,][]{Klypin-missing,Moore-substructure,Strigari-VL}. Unfortunately, it is difficult to constrain the mass function with anomalous flux ratios, because there is a degeneracy such that a small clump near a lensed image (in projected distance) can produce the same flux perturbation as a large clump farther away \citep[see][]{Dalal-substruc}.

One possible solution to this problem is to measure flux ratios at multiple wavelengths. Quasar emission regions are believed to have different sizes at different wavelengths. Roughly speaking, only clumps with Einstein radii (Einstein angles translated into units of length) larger than the size of the source can produce flux ratio anomalies \citep[e.g.,][]{Dobler-finite}; clumps with small Einstein radii act collectively as a smooth mass component. Since the Einstein radius depends on mass, the source size effectively translates into a mass threshold. The net effect is that multi-wavelength observations provide a way to study clumps with different mass ranges.\footnote{This is similar to the idea of ``chromatic'' microlensing, which is discussed by \citet{Kochanek-SCMA}.}  For example, continuum \citep{Wozniak-2237-continuum} and broad-line \citep{Richards-1004, Keeton-0924} optical emission regions in quasars can be perturbed by objects of stellar mass and larger. Radio emission regions, by contrast, are large enough that only objects with masses $\gtrsim 10^6 \Msun$ are important \citep{Dobler-finite}.  If a flux-ratio anomaly is observed at optical wavelengths but not at radio wavelengths, it is likely that microlensing by stars in the lens galaxy is the culprit.  To avoid contamination by microlensing, \citet{Dalal-substruc} focused on radio anomalies in their study. Infrared emission regions are intermediate in size, so observations in this band provide a way to probe objects with mass scales between stars and clumps. Recent mid-IR observations with the Spitzer \citep{Poindexter-1104-erratum, Poindexter-1104-IR} and Subaru \citep{Chiba-IR, 0414-2237-IR} telescopes have achieved the spatial resolution necessary for strong lensing studies. In particular, \citet{Chiba-IR} found evidence for subhalos of $\sim 10^4 \Msun$ in the lens system B1422+231, and possibly in PG 1115+080 as well.  In addition, \citet{0414-2237-IR} recently found evidence of a clump with mass $> 10^5 \Msun$ in the lens MG 0414+0534.  A related way to study the substructure mass function is to compare continuum and broad-line emission in optical lenses \citep{Moustakas-substructure}, which originate from the central accretion disk and an extended distribution of fast-moving clouds, respectively.  These two regions differ in length scale by a factor of $\sim$100, whence their utility for substructure lensing.  This technique has been employed to study the four-image lenses HE~0435$-$1223 \citep{Wisotzki-0435spec}, SDSS J0924+0219 \citep{Keeton-0924,Eigenbrod-0924}, SDSS J1004+4112 \citep{Richards-1004}, RX~J1131$-$1231 \citep{Sluse-1131spec}, and Q2237+0305 \citep{Metcalf-2237,Eigenbrod-2237}, along with several two-image lenses \citep{Burud-2149,Inada-0246,Inada-0806,Sluse-0806}

While the method of probing substructure with differential source size effects is promising, this approach does have some limitations. For one thing, the mass threshold defined by the source size and Einstein radius does not represent a sharp cutoff \citep[see][]{Dobler-finite}, so it is difficult to obtain tight constraints on the clump mass. In addition, it may not be possible to measure flux ratios at enough wavelengths to sample the mass function well. Finally, lensing constraints on the mass function will only be as accurate as the mapping of wavelength to source size.  While the model of \citet{Shakura-Sunyaev} has been quite successful in explaining the observed properties of accretion disks, and gives a relation between wavelength and source size, it does not apply to the full range of wavelengths relevant to strong lensing.

A second possible approach to constrain the substructure mass function is to use flux ratios in combination with other lens observables. Although they are smaller in amplitude than flux ratio perturbations, substructure effects on both image positions \citep{Chen-imgshift} and time delays \citep{Keeton-tdel} should be detectable. Combining different observables will be valuable because they depend on the lens potential in different ways: time delays, image positions, and flux ratios depend on the potential and its first and second derivatives, respectively. The different observables, in other words, contain different information about the substructure population. For example, \citet{Keeton-tdel} showed that time delays are sensitive to the slope and dynamic range of the substructure mass function, which suggests that precise time delay measurements may provide a way to constrain these properties. Before we can use this method, we need to determine whether observed time delays differ from the predictions for smooth mass models in a way that may indicate the presence of substructure. Developing a method to identify such ``time delay anomalies'' is the focus of this paper.

In particular, we wish to understand the range of time delays that can be produced by reasonable smooth models. We can then classify any outlier as ``anomalous'' and use it as evidence of complexity in the lens potential. We must be careful when interpreting anomalies, however, because dark matter substructure may not be the only relevant source of complexity in the lens galaxy. Stars also constitute complex structure that is important when interpreting flux ratios (due to microlensing), but they have essentially no effect on time delays \citep{Keeton-tdel}. This means that time delays should not depend on wavelength, so we do not need to distinguish between radio and optical time delays.\footnote{Strictly speaking, optical and radio emission come from different regions of a quasar, so the optical and radio image positions of a lens, and hence the corresponding time delays, need not be identical.  This effect is naturally taken into account by our criterion for matching observed time delays with simulated lenses (see \S\ref{sec:distributions}).}  Finally, extinction by dust in the lens galaxy can perturb flux ratios, but it will not have any effect on time delays, which are measured through flux variability alone and do not depend on color information.

The other main source of complexity we need to consider is the environment of the lens galaxy. Many lens galaxies lie in groups or clusters of galaxies \citep[e.g.,][]{Momcheva-env,Auger-env}. To lowest order, such an environment contributes a tidal shear to the lens potential \citep[e.g.,][]{Keeton-shear-ellip}, and we therefore include shear in our analysis, but there may be higher-order terms that are non-negligible. Extreme examples of this situation include the lens B1608+656, which has two galaxies inside the Einstein angle \citep{Koopmans-Fassnacht-1608, Koopmans-1608-model, Suyu-1608-model}, and the lens SDSS J1004+4112, which is produced by a cluster of galaxies \citep{Inada-1004,Oguri-1004}. In general, time delay anomalies in close pairs of images potentially provide the strongest evidence of dark matter substructure, because environmental effects are fairly large-scale and should not produce dramatic differences between images separated by a distance much less than the Einstein angle. With time delay anomalies in image pairs that have larger separations, by contrast, we will need to take more care to consider environmental effects as well as substructure.

Our approach follows that of \citet{Keeton-cusp, Keeton-fold}, who employed analytic flux-ratio relations that are generic for all lenses with fold and cusp configurations produced by smooth mass models. To be more specific, a fold lens contains a bright pair of images whose fluxes $F_A$ and $F_B$ should satisfy the relation
\begin{equation}\label{eqn:Rfold}
  \Rfold \equiv \frac{F_A - F_B}{F_A + F_B}
  \approx \Afold \, d_1,
\end{equation}
where $d_1$ is the image separation and $\Afold$ depends on properties of the lens potential. A cusp lens contains a triplet of bright images whose fluxes should satisfy the relation
\begin{equation}\label{eqn:Rcusp}
  \Rcusp \equiv \frac{F_A - F_B + F_C}{F_A + F_B + F_C}
  \approx \Acusp \, d_1^{2},
\end{equation}
where $d_1$ is the distance between the closest pair of images and $\Acusp$ depends on properties of the lens potential \citep[see, e.g.,][]{Congdon-tdel-pert}.  (Note that $\Rfold$ and $\Rcusp$ vanish in the limit $d_1\rightarrow 0$.)  If these relations are violated, we may conclude that the mass distribution of the lens galaxy cannot be smooth and must contain additional structure, most likely in the form of CDM substructure or a complex lens environment. Determining in practice that a given lens is anomalous requires some care.  Equations (\ref{eqn:Rfold}) and (\ref{eqn:Rcusp}) show that $\Rfold$ and $\Rcusp$ increase with the distance between the images, so a non-zero value for one of these quantities does not automatically imply a flux ratio anomaly.  Making such an identification would require knowledge of $\Afold$ and $\Acusp$, which depend on the exact form of the lens potential.  Since these quantities are not directly observable, \citet{Keeton-cusp, Keeton-fold} performed Monte Carlo simulations to determine distributions for $\Rfold$ and $\Rcusp$ at fixed $d_1$ using parameter values appropriate for a realistic population of lens galaxies.  With this information, it is possible to determine the probability that an observed lens is anomalous and hence contains small-scale structure.

\citet{Oguri_tdel} employed a method similar to that of \citet{Keeton-cusp, Keeton-fold} to study time delays, although his emphasis was different from ours.  While both studies are based on time delay distributions, Oguri's emphasis was on determining the Hubble constant, whereas we focus on small-scale structure in lens galaxies.  In fact, we explicitly remove the Hubble constant from our analysis by working with scaled time delays and time delay ratios.  Another difference is that we concentrate our attention mainly on cusp and fold lenses, which are most relevant to the questions we seek to answer.

This paper is organized as follows.  In \S\ref{sec:hcaus} we consider how the differential time delay depends on the parameters of the lens potential, and on the shape of the associated (astroid) caustic.  The results we present there motivate our Monte Carlo simulations, which we describe in \S\ref{sec:distributions}.  We apply our formalism to the 25 known four-image lenses in \S\ref{sec:results}.  Finally, we present our conclusions in \S\ref{sec:conclusions}.

\section{DEPENDENCE OF TIME DELAY ON LENS POTENTIAL AND POSITION ALONG CAUSTIC}
\label{sec:hcaus}

In \citet{Congdon-tdel-pert} we showed that the time delay between a close pair of images in a fold lens scales with the cube of the image separation, $d_1$, i.e., $\Delta\tau/\tau_0 \approx |h| d_1^3/2$ (see eq.\ [25] of \citealt{Congdon-tdel-pert}, and pp.\ 190--191 of \citealt{Schneider_lensing}), where $\tau_0$ is a cosmology-dependent scale factor (see eq.\ [\ref{eq:tau0}] below).  The coefficient $h$ comes from a Taylor expansion of the lens potential, $\psi$, and can be written as a particular third derivative: $h\equiv \psi_{222}/6$, where the subscript ``2'' indicates differentiation with respect to the coordinate in the image plane that corresponds to the direction perpendicular to the caustic in the source plane, and is evaluated at the point on the critical curve that serves as the coordinate origin (see \S3.2 of \citealt{Congdon-tdel-pert}). In this section we study how the time delay for a fold pair depends on the lens potential and the distance between the fold point and the nearest cusp point. This is equivalent to studying the variation of $h$ along the caustic, since, for fixed $d_1$, the time delay is given solely in terms of this coefficient. Because $h<0$, we find it more convenient to work with its absolute value.

Most lens galaxies are of early type and have density profiles close to isothermal, i.e., the three dimensional density scales with radius as $\rho \propto r^{-2}$ \citep[e.g.,][]{Treu-lensing-dynamics,Rusin-field-ellip,Treu-SLACS-II}, so we compute $|h|$ for a singular isothermal ellipsoid (SIE) lens. To determine an appropriate value for the ellipticity parameter $e \equiv 1 - q$, where $q$ is the minor-to-major axis ratio, we turn to the observed galaxy samples of \citet{Bender-ellip}, \citet{Jorgensen-ellip}, and \citet{Saglia-ellip}. These samples have mean ellipticities and dispersions of $(\bar e, \sigma_e) = (0.28, 0.15), (0.31, 0.18)$, and $(0.30, 0.16)$ respectively.  Note that these values measure the distribution of light rather than mass, so it is possible that the dark matter halo in which the galaxy presumably resides is rounder or flatter than the observed isophotes.

\begin{figure}[t]
\begin{center}
\includegraphics[width=0.32\textwidth]{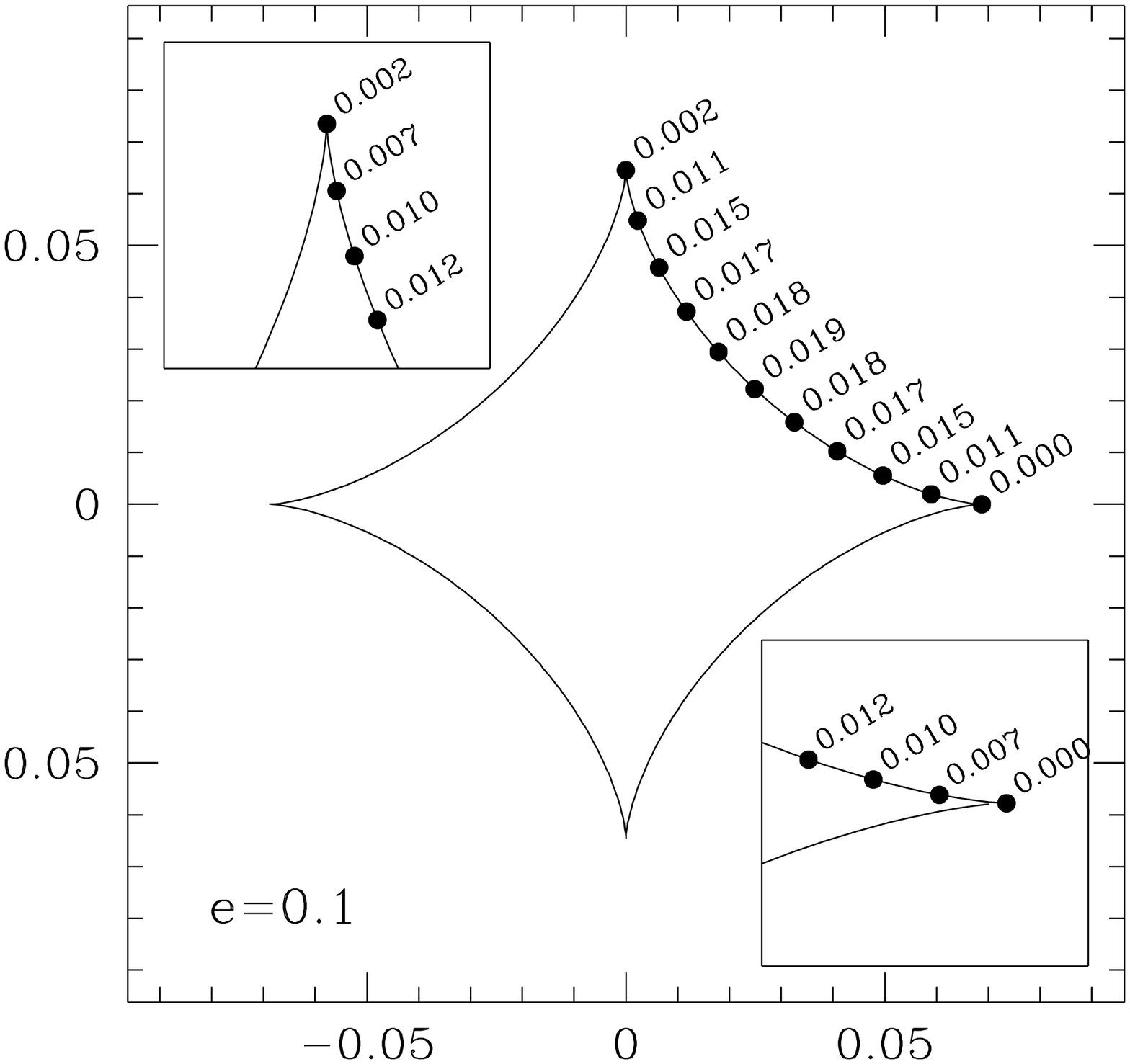}
\includegraphics[width=0.32\textwidth]{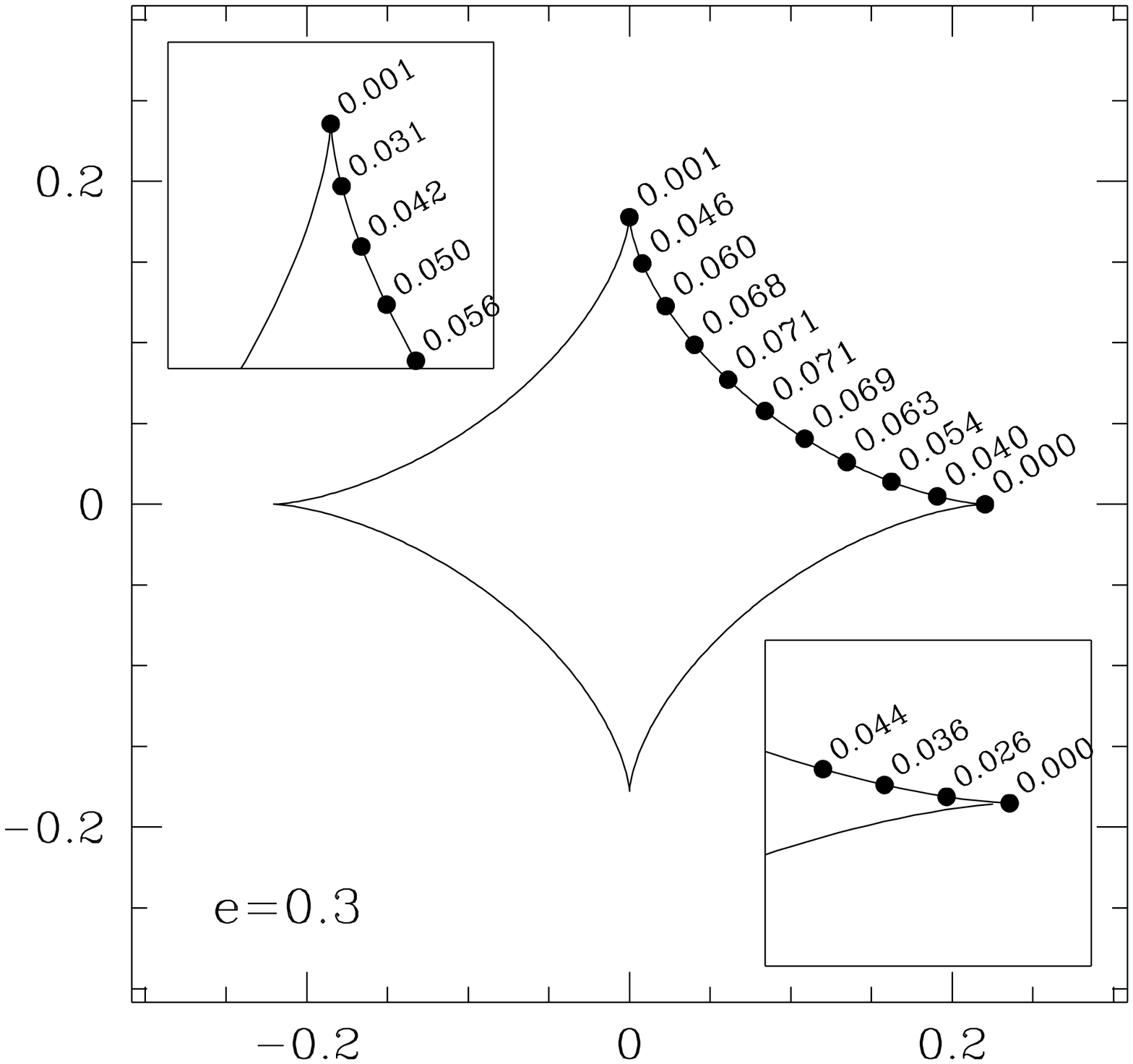}
\includegraphics[width=0.32\textwidth]{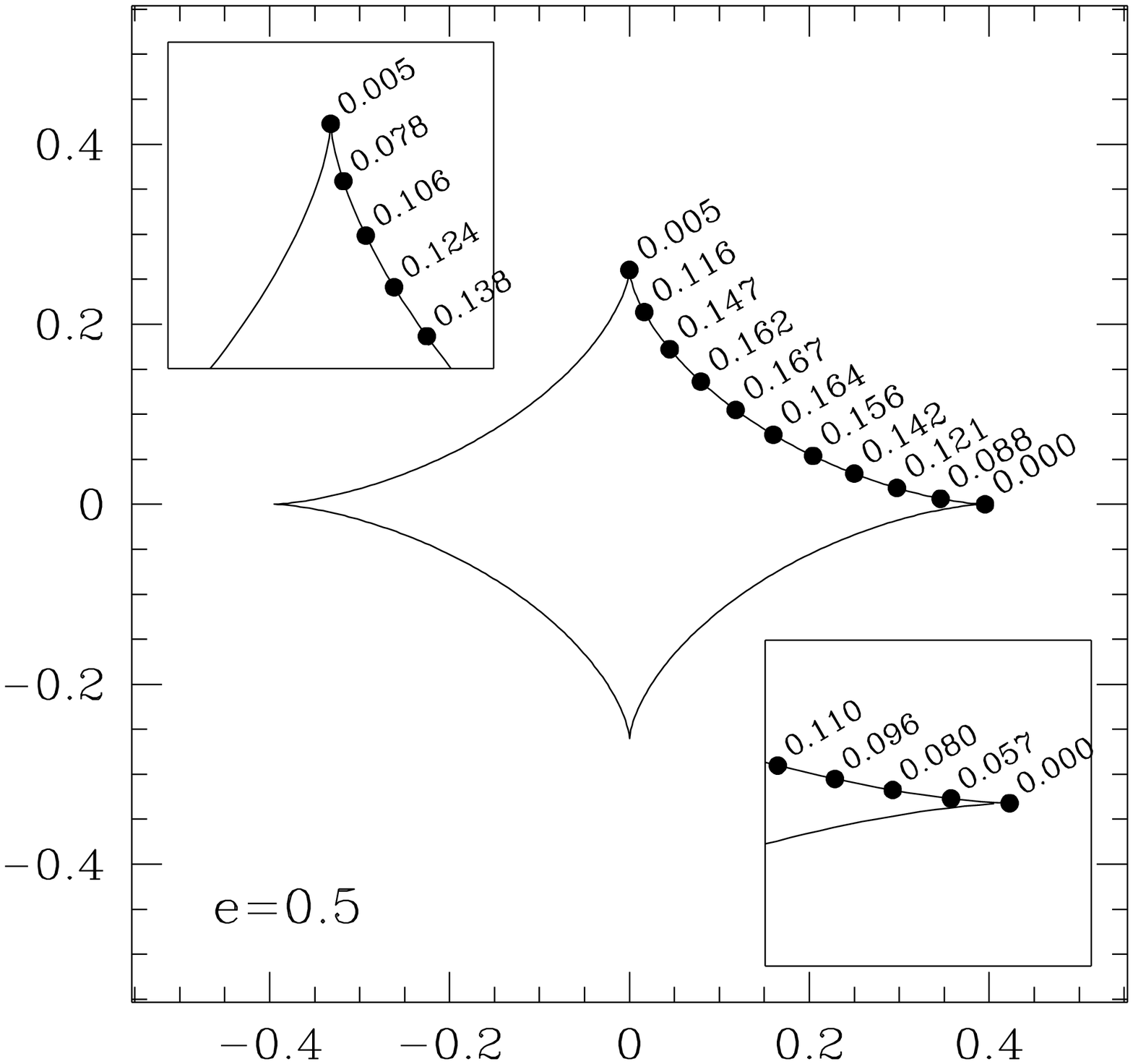}
\caption{
Values of the Taylor series coefficient $|h|$ at various points on the caustic for a singular isothermal ellipsoid lens with different values of the ellipticity.  Insets show close-up views of the upper and right-hand cusp points. The coefficient $h$ should vanish at a cusp point \citep[see][]{Congdon-tdel-pert}, but for computational reasons we do not necessarily have points that lie precisely at the cusps.  The axes are in units of the Einstein angle.
\label{fig:hcaus}}
\end{center}
\end{figure}

Figure \ref{fig:hcaus} shows $|h|$ at various points on the caustic, for ellipticities of 0.1, 0.3, and 0.5.  We see that $|h|$ remains roughly constant along the caustic for points away from the cusps. This suggests that lenses whose fold pairs have comparable separations will have similar time delays as well, at least for galaxies with similar ellipticities. Testing this prediction will require large samples of fold lenses, for which both the differential time delay between the fold pair and the ellipticity of the lens galaxy are known. For points near a cusp, $|h|$ decreases rapidly and vanishes right at the cusp point (as it must; see \S2 of \citealt{Congdon-tdel-pert}). Finally, we see that $|h|$ depends on the size of the caustic. It is not yet clear whether this merely reflects a simple correlation between $|h|$ and $e$ or is indicative of a more subtle relationship between caustic size and the time delay for a fold pair.

In addition to ellipticity, \citet{Bender-ellip} and \citet{Saglia-ellip} find that many early-type galaxies have small departures from elliptical symmetry. ``Disky'' or ``boxy'' isophotes can be represented by an $m=4$ multipole mode, with coefficient $a_4 > 0$ for disky isophotes and $a_4 < 0$ for boxy isophotes. In terms of the (two-dimensional) lens potential, the $m=4$ mode for an isothermal galaxy can be written as (e.g., \citealp{Keeton-gravlens})
\begin{equation}
\psi(r,\theta)=-A_4 r \cos{4\theta} \,.
\label{eqn:m4}
\end{equation}
The coefficient $a_4$ is defined in terms of surface brightness rather than potential, so we must convert from $A_4$ to $a_4$:
\begin{equation}
a_4 = 15A_4 \sqrt{1-\epsilon} \,,
\label{eqn:m4}
\end{equation}
where $\epsilon \equiv (1 - q^2)/(1 + q^2)$. For typical values of $a_4 = -0.01$, 0, 0.01, and 0.02 (e.g., \citealp{Bender-ellip}) and ellipticity $e=0.3$, we find maximum $|h|$-values of 0.07, 0.07, 0.08, and 0.09, respectively. We therefore conclude that $m=4$ multipole terms have fairly small effects on the time delay for a fold pair, regardless of whether the isophote is disky or boxy.  We nevertheless include $m=4$ modes in our simulations (\S\ref{sec:distributions}) given that this is both simple and observationally motivated.

Since many lens galaxies lie within groups or clusters \citep[e.g.,][]{Momcheva-env,Auger-env}, a nonzero tidal shear is common. Using numerical simulations and semianalytic models of galaxy formation, \citet{Holder-shear} find that shear can be described by a lognormal distribution with median $\gamma = 0.05$ and dispersion $\sigma_\gamma = 0.2$ dex.\footnote{Following \citet{Keeton-fold}, we do not quote the final shear distribution of \citet{Holder-shear}, but rather the raw distribution before effects such as magnification bias are applied.} For such shear amplitudes, the caustic structure is qualitatively similar to what is seen in Figure \ref{fig:hcaus}, so we do not show it explicitly. We return to shear when we consider realistic lens populations in the following section.

\section{CONSTRUCTING TIME DELAY DISTRIBUTIONS FOR A REALISTIC LENS POPULATION}
\label{sec:distributions}

For any given lens galaxy, we could estimate $h$ (or more generally the smooth model time delays) directly, by fitting a lens model and inferring the lens potential.  However, we would like to avoid any dependence on models and modeling as much as possible. We therefore elect to examine the full probability distributions for time delays given a realistic population of lens galaxies, using Monte Carlo methods to construct the time delay distributions.

We follow the approach of \citet{Keeton-cusp, Keeton-fold} and create lens galaxies with model parameters drawn from the galaxy samples of \citet{Bender-ellip}, \citet{Jorgensen-ellip}, and \citet{Saglia-ellip}. Although the mean ellipticities and dispersions are similar for the three samples, the underlying galaxy populations are not the same. \citet{Jorgensen-ellip} and \citet{Saglia-ellip} include galaxies within clusters, while \citet{Bender-ellip} use bright, nearby galaxies to construct their sample. In addition, \citet{Bender-ellip} and \citet{Saglia-ellip} report $a_4$ measurements, while \citet{Jorgensen-ellip} do not. In order to determine the extent to which our conclusions depend on the input data, we carry out the numerical method described below separately for each of the three galaxy samples.  To model the environment of a lens galaxy, we add tidal shear based on the simulations of \citet{Holder-shear} that we discussed at the end of the previous section.  For the sample of \citet{Jorgensen-ellip} we pick 2000 ellipticities from their observed distribution, and assign a random shear to each.  For the samples of \citet{Bender-ellip} and \citet{Saglia-ellip}, we use the actual $(e,a_4)$ pairs for the observed galaxies (87 in the \citealt{Bender-ellip} sample, and 54 in the \citealt{Saglia-ellip} sample) in order to retain any correlation between the parameters; and for each galaxy we use 100 different realizations of the shear.

For each model lens potential, we use an updated version of the {\em GRAVLENS} software\footnote{See http://redfive.rutgers.edu/$\sim$keeton/gravlens} \citep{Keeton-gravlens} to solve the lens equation numerically and obtain the image positions and time delays for a large set of random source positions that are uniformly distributed in the four-image region.  Using a uniform distribution has several consequences.  First, it means that each model lens potential is automatically weighted by its four-image cross section, which seems like the proper statistical approach.\footnote{The mass of the lens galaxy factors out when we work with scaled time delays and time delay ratios, but the angular structure of the lens potential remains important.}  Second, it means that we neglect magnification bias.  While lensing magnification bias is quite important when comparing statistical samples of four-image and two-image lenses, it is less dramatic within a sample of four-image lenses, and still less so within the subset of four-image lenses that have a particular image configuration.  Generally speaking, including magnification bias would give more weight to sources that lie closer to the lensing caustic, which tend to produce shorter time delays, so it would shift the time delay distributions we derive to somewhat shorter values.  The effect is not strong, however, and we checked that it does not change any of our conclusions about time delay anomalies.  All told, our simulations based on the galaxy samples from \citet{Bender-ellip}, \citet{Jorgensen-ellip}, and \citet{Saglia-ellip} contain 1,267,555, 2,205,515, and 851,261 mock four-image lens systems, respectively.

The time delay of an image at angular position $\btheta$ relative to an unlensed light ray from the true source with position $\bbeta$ is given by
\begin{equation} \label{eq:tau}
  \tau(\btheta) = {\tau_0} \left[ \frac{1}{2} \left|\btheta-\bbeta\right|^2
    - \psi(\btheta) \right] ,
\end{equation}
where the time scale is
\begin{equation} \label{eq:tau0}
  \tau_0 = \frac{1+z_L}{c} \frac{D_L D_S}{D_{LS}} .
\end{equation}
Here $D_L, D_S$, and $D_{LS}$ are the angular-diameter distances from the observer to lens, observer to source, and lens to source, respectively. For our mock lenses we focus on the dimensionless, scaled time delay
\begin{equation} \label{eq:tauhat}
  \hat\tau \equiv \frac{\tau}{\tau_0 \theta_E^2}\ ,
\end{equation}
where $\theta_E$ is the Einstein angle.\footnote{The Einstein angles we use come from lens modeling, either by ourselves or others (see Table \ref{tab:lens-info}). Although there is no unique definition for the Einstein angle of an elliptical lens, different definitions result in values consistent to within a factor of order unity. In practice, such concerns are rendered moot by the 5\% astrometric uncertainties we allow for when comparing observed lenses with our mock catalog (see text for details).} The advantage of working with the scaled time delay is that our analysis of the mock lenses is independent of cosmology, and of the lens and source redshifts. It does mean that we need to convert all observed time delays from physical to dimensionless units before we can compare them to our mock lens catalog (see \S \ref{sec:results}).  For convenience, we quote distances (e.g., $d_1$ and $d_2$ below) in units of the Einstein angle as well.

Our analytic time delay relation (eq.~25 of \citealt{Congdon-tdel-pert}) specifically applies to fold image pairs, so they are the most obvious targets for anomaly searches. However, we do not actually use the analytic relation to make predictions for our mock lenses, so it is reasonable to consider non-fold image pairs, as well as cusp and cross lenses, in our comparisons between observed lenses and Monte Carlo simulations. There are six distinct image pairs in a four-image lens. We are interested in the subset of pairs that contain one image of positive parity (which lies at a minimum of the time delay surface) and one of negative parity (which lies at a saddlepoint). In this way, we consider only adjacent images.  The effect is to make our analysis local\footnote{It would perhaps be better to say that our analysis is quasi-local. It is indeed local for close pairs of images (as in fold pairs and cusp triplets), but it is not strictly local for image pairs whose image separation $d_1$ approaches order unity.} rather than global, which is appropriate for studying small-scale structure. There are four such mixed-parity image pairs in each four-image lens.

We characterize each image pair by the separation between the two images, $d_1$. We also use the distance to the next-nearest image, $d_2$,\footnote{To obtain $d_2$ we consider the distance between each image of a given pair and its next-nearest image of opposite parity.  We define $d_2$ to be the smaller of these two distances.} because it helps specify the lens morphology: a fold lens has $d_1 \ll d_2 \sim 1$; a cusp lens has $d_1 \sim d_2 \ll 1$; a cross lens has $d_1 \sim d_2 \sim 1$. Another reason to use $d_2$ is that it allows us to take ratios of time delays (see below). While $d_1$ and $d_2$ technically define an image triplet, we will use the term ``pair'' to refer to the two images defined by $d_1$, since this is the image pair of primary interest.

\begin{figure*}[t]
\begin{center}
\includegraphics[width=0.8\textwidth]{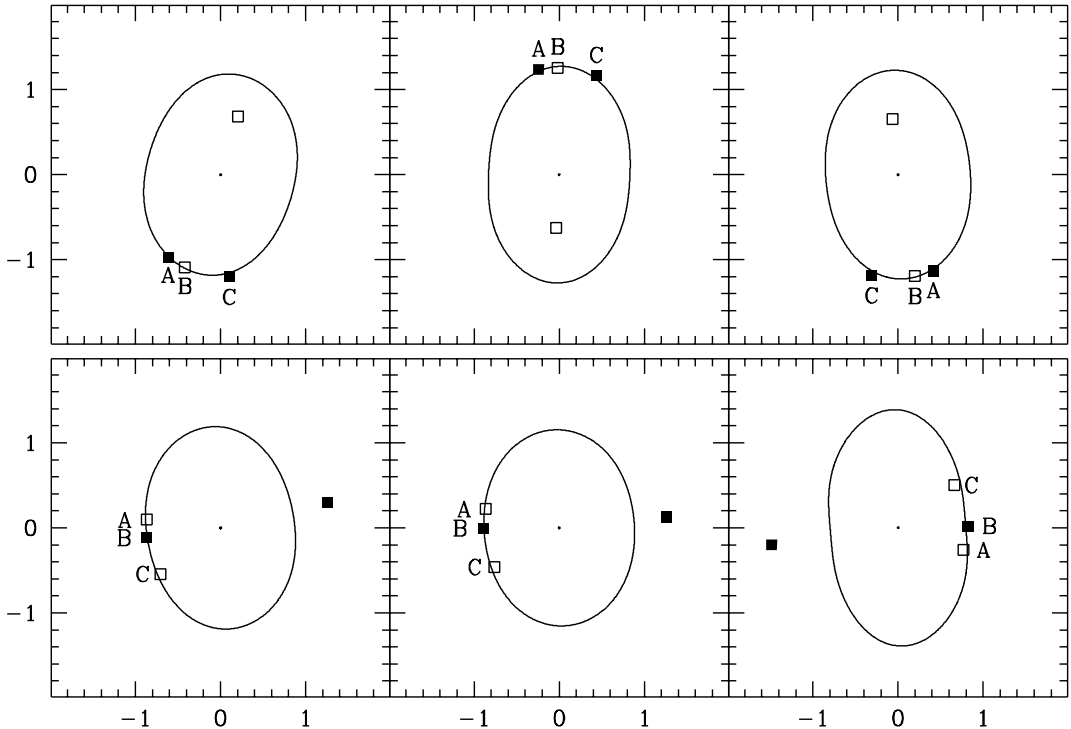}
\caption{
Examples of mock lenses that potentially match the cusp lens B2045+265.  Specifically, the distances $d_1$ (between images A and B) and $d_2$ (between images B and C) match the corresponding distances in 2045 within a tolerance of $\pm$0.05. Filled squares denote minima, while open squares denote saddles. The lensing critical curves are shown, and the axes are labeled in units of the Einstein angle, $\theta_E$. The major axis of the lens galaxy is oriented vertically, so the mock lenses in the top row are long-axis cusps, while the mock lenses in the bottom row are short-axis cusps.  When the parity cut is applied, only the top panels match 2045.
\label{fig:sampcusp}}
\end{center}
\end{figure*}

We seek to understand the full range of time delays that can be produced by smooth lens models that are consistent with the positions of an observed image pair. We consider a mock lens image pair to ``match'' an observed pair if two conditions are satisfied.  First, we require that the $d_1$ and $d_2$ values of the mock and observed pairs agree within a tolerance of $\pm$0.05. This criterion is conservative in the sense that we are using minimal information from quantities that are observable and local; using additional information would only narrow the range of smooth models that are considered to be consistent with an observed image pair. The distance-based matching criterion does have one limitation, which is illustrated in Figure \ref{fig:sampcusp}: it may allow not only mock pairs with the same parities as the observed images, but also pairs with the opposite parities.  We could avoid this problem by adding information about the fourth image. We do not want to do that, however, because the fourth image is always ``far'' from the pair in question (i.e., 1--2 Einstein angles away); including it would make our analysis more sensitive to global properties of the lens potential and countermand our goal of using a local analysis to search for substructure. Instead, we add the image parities to the matching criterion.\footnote{The parity cut was not used by \citet{Keeton-cusp,Keeton-fold}, but we introduce it here in part because \citet{Keeton-tdel} note the importance of parity in the context of time delays.}  Specifically, we insist that the parities of the three images defined by $d_1$ and $d_2$ are the same for an observed lens and its simulated counterpart.  This is straightfoward to do since the image parities are known for the mock lenses, and they can be determined unambiguously for observed lenses (by measuring time delays or just analyzing the image configuration; see, e.g., \citealt{Saha-img-order}). Since the two images in each pair we consider have opposite parities (by construction), the parity cut represents only one additional bit of information.

For each image pair we compute the differential scaled time delay between the two images, which we call $\Delta t_1$. For the second differential time delay we use the image pair characterized by the distance $d_2$, and we call this $\Delta t_2$. We then adopt the following sign convention. By construction, any image pair we consider will consist of a minimum image (positive parity, labeled $M_1$) and a saddle image (negative parity, labeled $S_1$). If the next-nearest image is a minimum (labeled $M_2$), we have a triplet consisting of a saddle flanked by two minima. We define the two differential time delays to be
\begin{equation} \label{eq:minsadmin}
  \Delta t_1 = \hat\tau(M_1) - \hat\tau(S_1) \quad {\rm and} \quad \Delta t_2 = \hat\tau(M_2) - \hat\tau(S_1) .
\end{equation}
Since saddles have larger time delays than minima, both $\Delta t_1$ and $\Delta t_2$ are negative. In the case that the next-nearest image is a saddle (labeled $S_2$), we have a triplet consisting of a minimum flanked by two saddles. We then define the differential time delays to be
\begin{equation} \label{eq:sadminsad}
  \Delta t_1 = \hat\tau(S_1) - \hat\tau(M_1) \quad {\rm and} \quad \Delta t_2 = \hat\tau(S_2) - \hat\tau(M_1) .
\end{equation}
Now both $\Delta t_1$ and $\Delta t_2$ are positive.

It is also useful to compute the ratio of the differential time delays for two image pairs. Time delay ratios are attractive because they do not depend on cosmology ($\tau_0$ factors out), and they are largely immune to the radial-profile degeneracy \citep[e.g.,][]{Kochanek-tdel,Keeton-tdel}.  Our sign convention ensures that $\Delta t_1 / \Delta t_2$ is always positive since $\Delta t_1$ and $\Delta t_2$ have the same sign.

To reprise, we define the two time delays relative to the ``middle'' of the three images we use to determine $d_1$ and $d_2$. This means the scaled time delays can be either positive or negative, but the time delay ratio is always positive. Note that regardless of the sign, when labeling image pairs we always list the minimum first.

From the set of mock lenses that match an observed image pair, we construct histograms of the scaled time delay and the time delay ratio. These represent the range of values that can be produced by realistic smooth mass distributions. We can then use these predicted distributions to assess whether observed time delays are or are not consistent with lensing by a smooth mass distribution.

\section{RESULTS FOR OBSERVED LENSES}
\label{sec:results}

\begin{deluxetable*}{lrrrr}
\tablewidth{0pt} \tabletypesize{\small} \tablecaption{Data for four-image lenses \label{tab:lens-info}} \tablehead{
Lens Name & $z_S$ & $z_L$ & $\theta_E$ (arcsec) & $\tau_0$ (days arcsec$^{-2}$)
}
\startdata
B0128+437   &   3.12    &   ?   &   0.20    &   ?   \\
HE 0230$-$2130    &   2.16    &   0.52    &   0.82    &   85.7    \\
MG 0414+0534    &   2.64    &   0.96    &   1.08    &   193.2   \\
B0712+472   &   1.34    &   0.41    &   0.68    &   72.9    \\
HS 0810+2554    &    1.50    &    ?    &    0.51    &    ?    \\
SDSS 0924+0219  &   1.52    &   0.39    &   0.87    &   64.8    \\
SDSS J1004+4112  &   1.73    &   0.68    &   6.91    &   140.7   \\
PG 1115+080 &   1.74    &   0.31    &   1.03    &   46.5    \\
SDSS J1330+1810    &    1.39    &    0.37    &    0.97    &    62.8    \\
B1555+375   &   ?   &   ?   &   0.24    &   ?   \\
B1608+656   &   1.39    &   0.63    &   0.77    &   143.7   \\
B1933+503   &   2.63    &   0.76    &   0.49    &   135.8   \\
WFI 2026$-$4536   &   2.23    &   ?   &   0.65    &   ?   \\
WFI 2033$-$4723   &   1.66    &   0.66    &   1.06    &   137.6   \\
\hline
RX J0911+0551   &   2.80    &   0.77    &   0.95    &   134.8   \\
RX J1131$-$1231   &   0.66    &   0.30    &   1.81    &   65.0    \\
SDSS J1251+2935   &   0.80    &   0.41    &   0.88    &   102.8   \\
B1422+231   &   3.62    &   0.34    &   0.76    &   46.1    \\
B2045+265   &   1.28    &   0.87    &   1.13    &   342.1   \\
\hline
HE 0435$-$1223    &   1.69    &   0.46    &   1.18    &   75.0    \\
HST 12531$-$2914  &   ?   &   0.69    &   0.55    &   ?   \\
HST 14113+5211   &   2.81    &   0.46    &   0.83    &   68.8    \\
H1413+117   &   2.55    &   ?   &   0.56    &   ?   \\
HST 14176+5226  &   3.40    &   0.81    &   1.33    &   135.7   \\
Q2237+030   &   1.69    &   0.04    &   0.85    &   4.9 \\
\enddata
\tablecomments{
Data for the known four-image lenses with point-like images. All data are given by \citet{Oguri_tdel} and the CASTLES website (http://www.cfa.harvard.edu/castles/), except for SDSS J1330+1810 \citep{Oguri-1330-discovery} and SDSS J1251+2935 \citep{Kayo-1251}. Question marks indicate quantities for which no measured value is available. Einstein angles $\theta_E$ are computed from lens models, either our own or others' (CASTLES, \citealt{Keeton-cusp, Keeton-fold, Kayo-1251, Oguri-1330-discovery}). The timescale $\tau_0$ depends on $z_L$ and $z_S$. We assume cosmological parameters $\Omega_M = 0.3$, $\Omega_\Lambda = 0.7$, and $H_0 = 70$ km s$^{-1}$ Mpc$^{-1}$. The table is divided into three sections: fold lenses (top), cusp lenses (middle), and cross lenses (bottom).
}
\end{deluxetable*}

The number of observed four-image lenses has been steadily increasing in recent years.  We focus on the 25 lens systems for which the lensed images are point-like (see Table \ref{tab:lens-info}). This restriction prevents us from using the system Q0047$-$2808 \citep{Warren96,Warren99}, as well as lenses from the SLACS survey\footnote{See http://www.slacs.org/} \citep{Bolton_SLACS1,Bolton-SLACS-V}.  However, this condition is necessary to ensure that the lensed source is compact, so that the variability timescale is short enough to make time delay measurements practical.

As noted above (see eq. [\ref{eq:tauhat}]), we compute scaled time delays for our mock lenses, so we must scale each observed time delay by $\tau_0 \theta_E^2$ in order to compare it with the mock lens sample. The Einstein angles and lens and source redshifts are listed in Table \ref{tab:lens-info}. We compute the angular diameter distances in $\tau_0$ assuming cosmological parameters $\Omega_M = 0.3$, $\Omega_\Lambda = 0.7$, and $H_0 = 70$ km s$^{-1}$ Mpc$^{-1}$. The benefit of working with scaled time delays in our mock lens analysis is that we can present results for all observed lenses, regardless of whether any of their time delays are currently known. As more time delays are measured, it will be a simple matter to compare them with our compilation of predicted scaled time delays. In addition, it is possible to consider different cosmological models simply by recomputing the scale factor $\tau_0$ and retranslating between observed and scaled time delays.

\subsection{Time Delay Histograms}
\label{sub:tdel-hist}

\begin{figure*}[h]
\begin{center}
\includegraphics[width=1.0\textwidth]{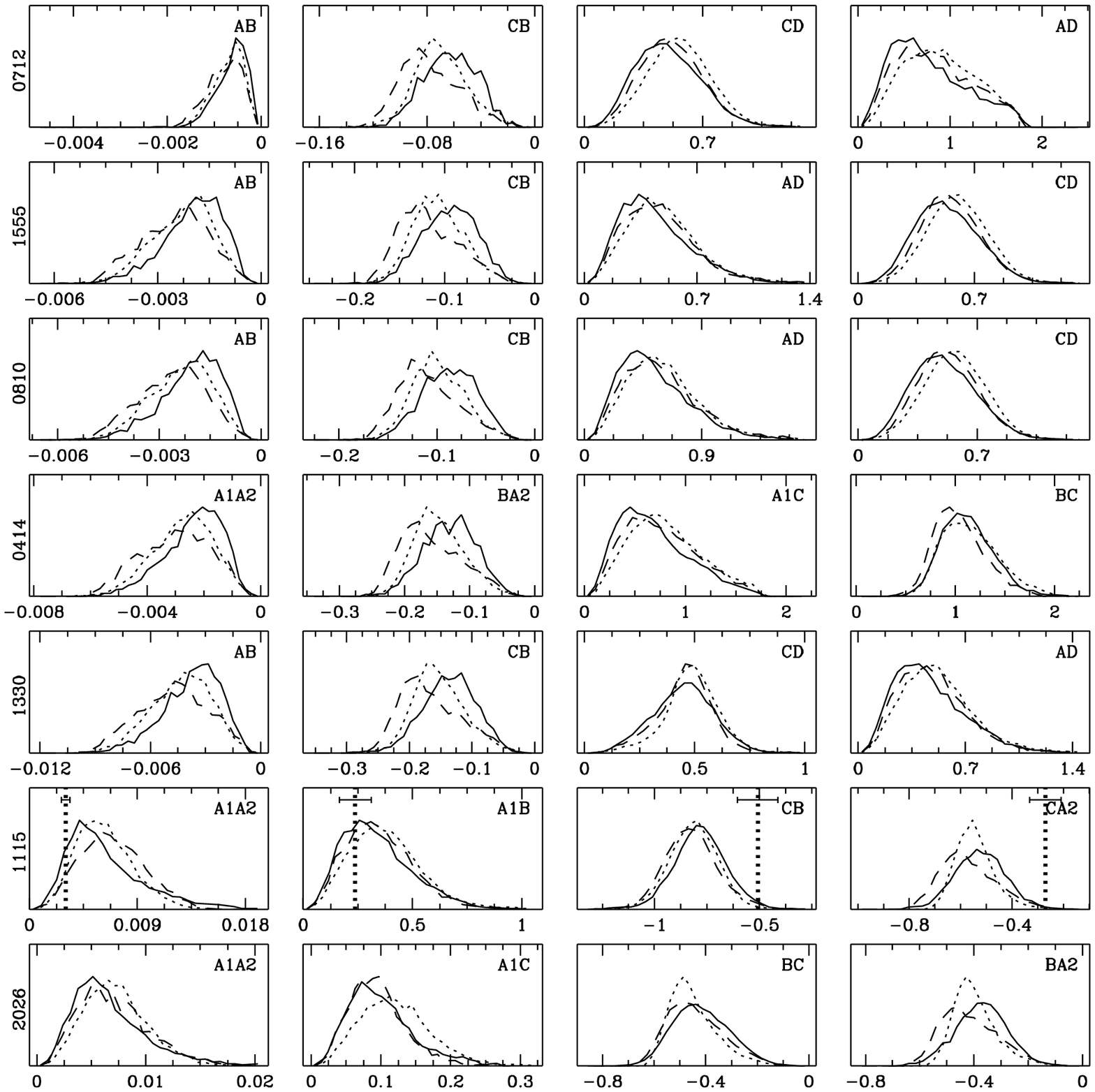}
\caption{
Time delay histograms for the known fold lenses. The horizontal axes show the scaled time delay in units of $\tau_0 \theta_E^2$; the range is chosen to span three standard deviations above and below the mean of the {\it predicted} time-delay distribution, and further expanded if necessary to encompass the observed value. The vertical axes are in arbitrary units, with each panel scaled to the maximum value of its three histograms. From top to bottom, lenses are arranged in order of increasing $d_1^{\,*}$; the abbreviated lens name appears at the far left. (See Table \ref{tab:lens-info} for the full names.) From left to right, the panels correspond to image pairs with increasing values of $d_1$. The solid, dotted, and dashed curves show histograms corresponding to the data of \citet{Bender-ellip}, \citet{Jorgensen-ellip}, and \citet{Saglia-ellip}, respectively. For image pairs with observed time delays, vertical dashed lines show the measured values.  The errorbars show 3$\sigma$ measurement uncertainties (see Table \ref{tab:pval-scaled}).  In cases where measurement uncertainty formally allows for time delays whose signs are disallowed by the parity cut (see \S\ref{sec:distributions}), errorbars are truncated at vanishing abscissa.
\label{fig:tdel-fold}}
\end{center}
\end{figure*}

\begin{figure*}[h]
\begin{center}
\includegraphics[width=1.0\textwidth]{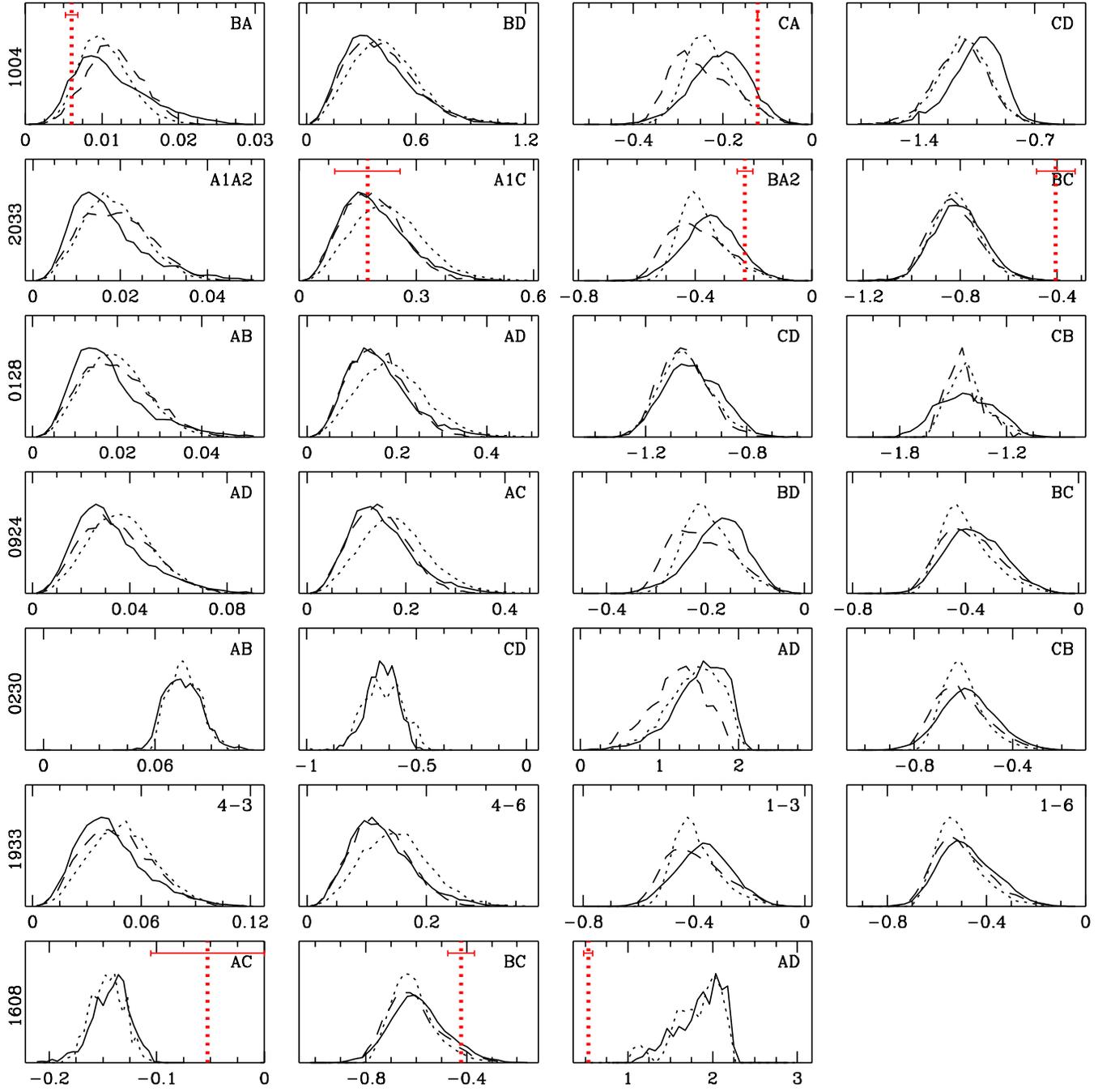}
\caption{
Time delay histograms for fold lenses (continued).
\label{fig:tdel-fold2}}
\end{center}
\end{figure*}

\begin{figure*}[h]
\begin{center}
\includegraphics[width=1.0\textwidth]{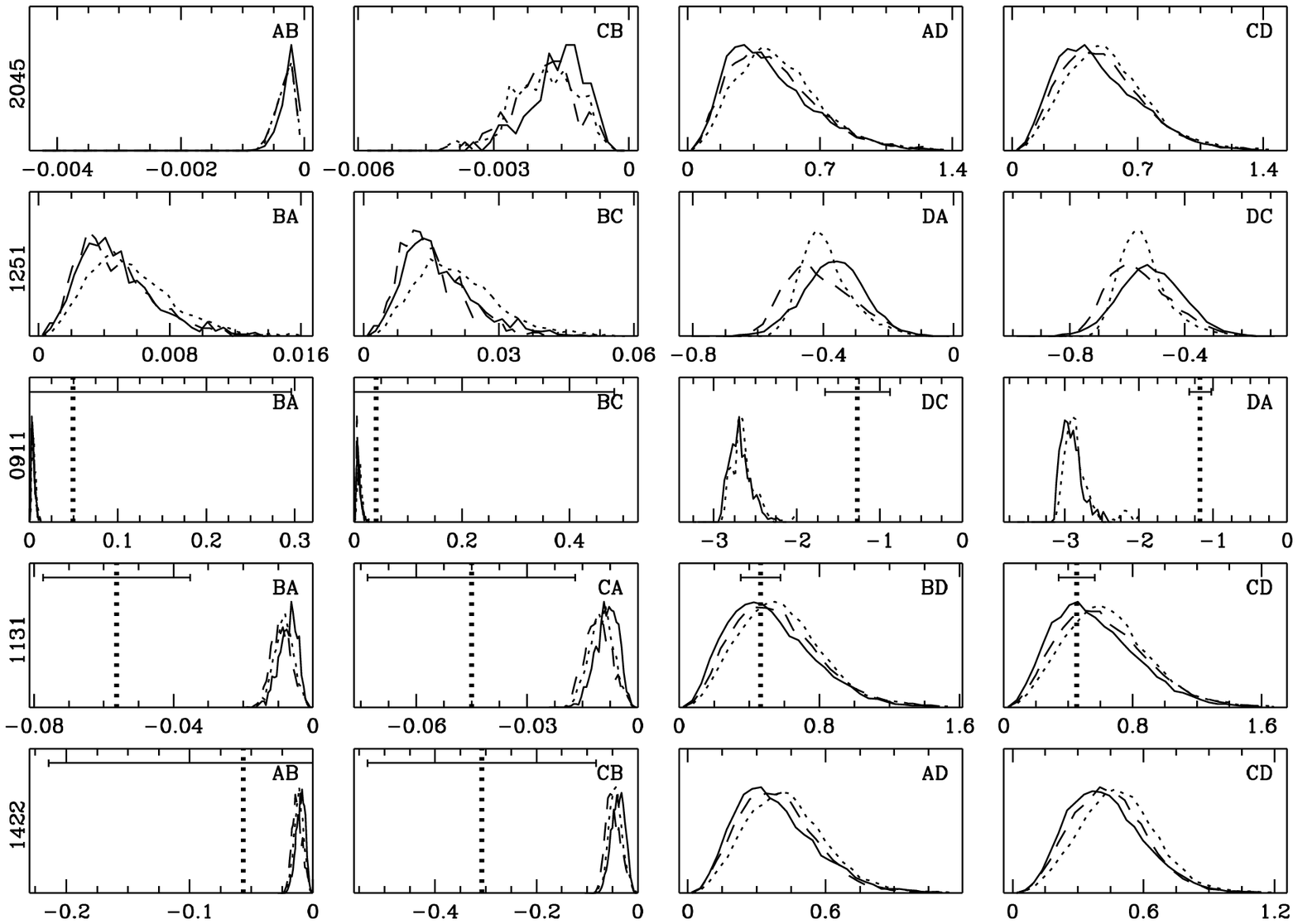}
\caption{
Same as Figure \ref{fig:tdel-fold}, but for the known cusp lenses.
\label{fig:tdel-cusp}}
\end{center}
\end{figure*}

\begin{figure*}[h]
\begin{center}
\includegraphics[width=1.0\textwidth]{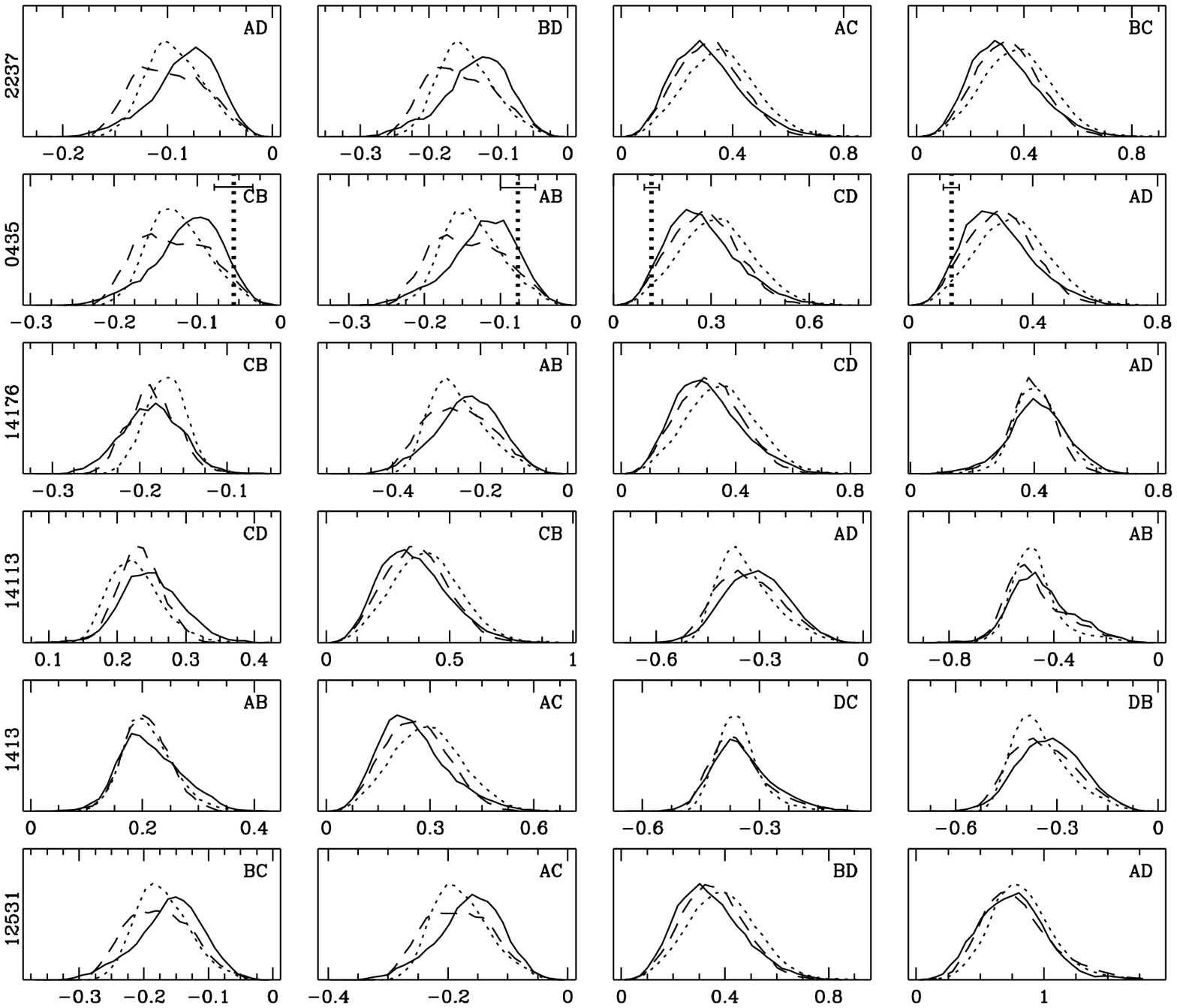}
\caption{
Same as Figure \ref{fig:tdel-fold}, but for the known cross lenses.
\label{fig:tdel-cross}}
\end{center}
\end{figure*}

\begin{figure*}[h]
\begin{center}
\includegraphics[width=1.0\textwidth]{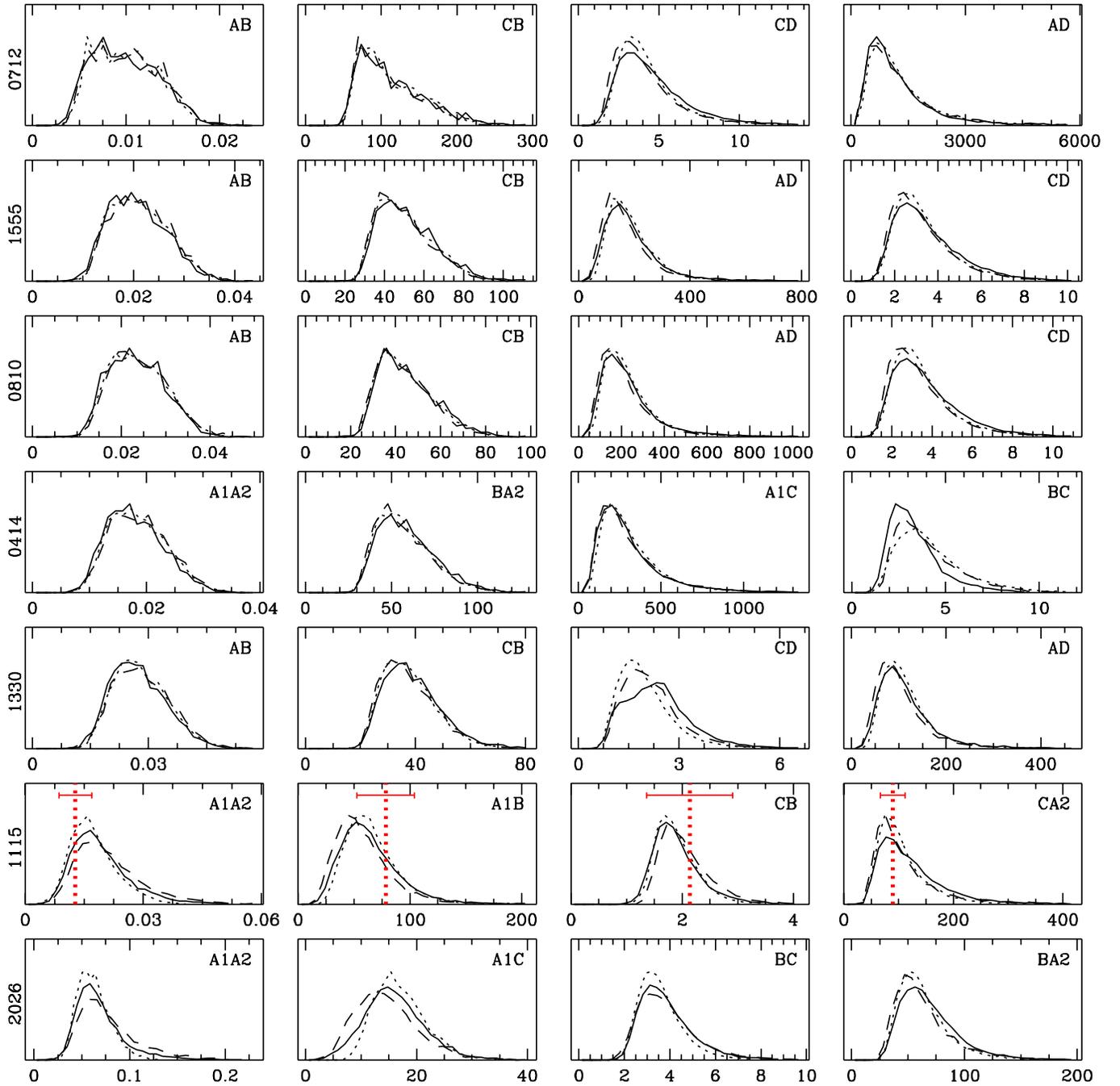}
\caption{
Histograms of time delay ratios for the known fold lenses. The horizontal axes show the (dimensionless) time delay ratio. The vertical axes are in arbitrary units, with each panel scaled to the maximum value of its three histograms. From top to bottom, lenses are arranged in order of increasing $d_1^{\,*}$; the abbreviated lens name appears at far left. (See Table \ref{tab:lens-info} for the full names.) From left to right, the panels correspond to image pairs with increasing values of $d_1$. The solid, dotted, and dashed curves show histograms corresponding to the data of \citet{Bender-ellip}, \citet{Jorgensen-ellip}, and \citet{Saglia-ellip}, respectively. For image pairs where it is possible to construct time delay ratios from observational data, vertical dashed lines show these values.  The errorbars show 3$\sigma$ measurement uncertainties (see Table \ref{tab:pval-rat}).  Truncated errorbars have the same meaning as in Figure \ref{fig:tdel-fold}.
\label{fig:trat-fold}}
\end{center}
\end{figure*}

\begin{figure*}[h]
\begin{center}
\includegraphics[width=1.0\textwidth]{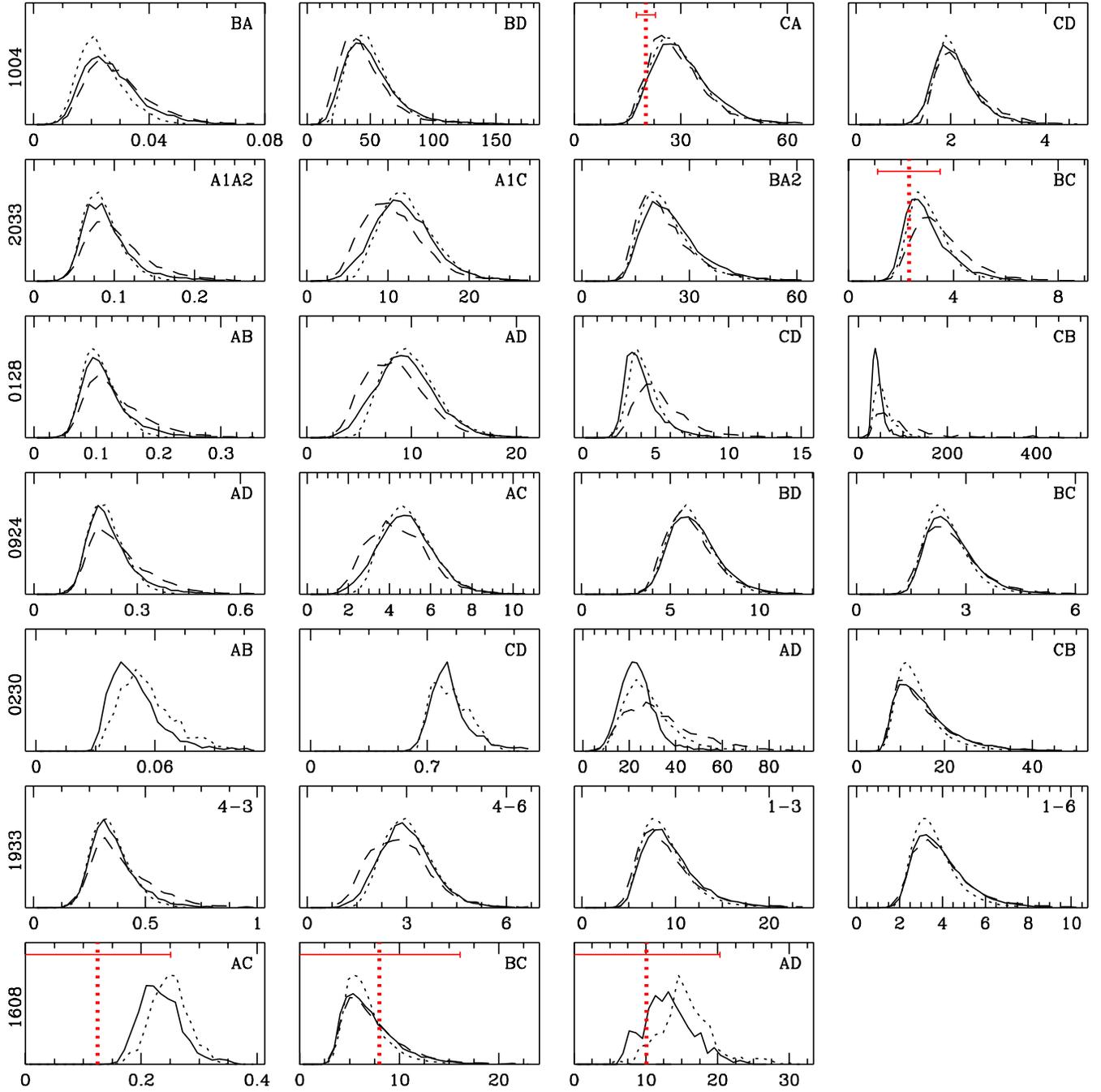}
\caption{Histograms of time delay ratios for fold lenses (continued).}
\label{fig:trat-fold2}
\end{center}
\end{figure*}

\begin{figure*}[h]
\begin{center}
\includegraphics[width=1.0\textwidth]{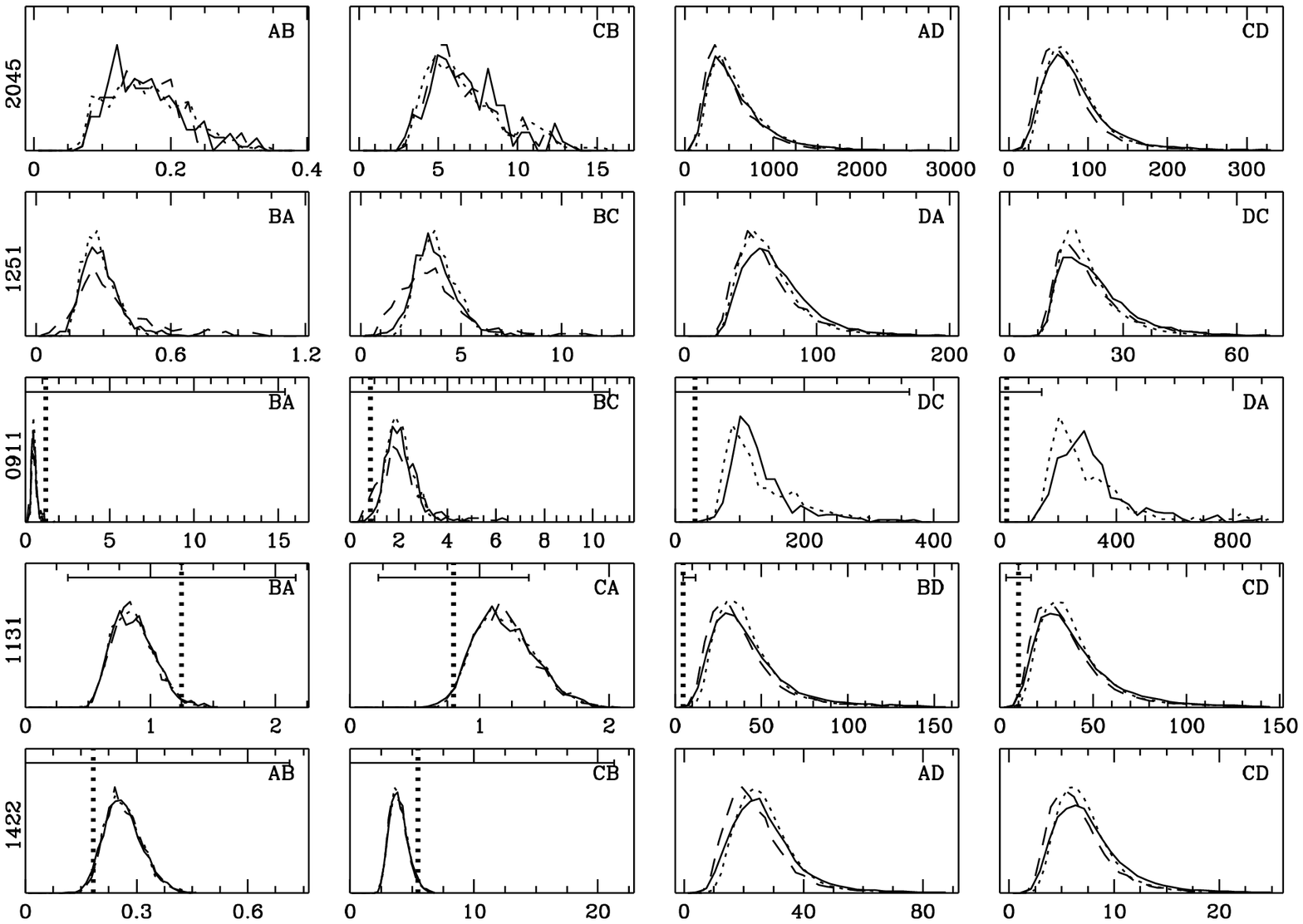}
\caption{
Same as Figure \ref{fig:trat-fold}, but for the known cusp lenses.
\label{fig:trat-cusp}}
\end{center}
\end{figure*}

\begin{figure*}[h]
\begin{center}
\includegraphics[width=1.0\textwidth]{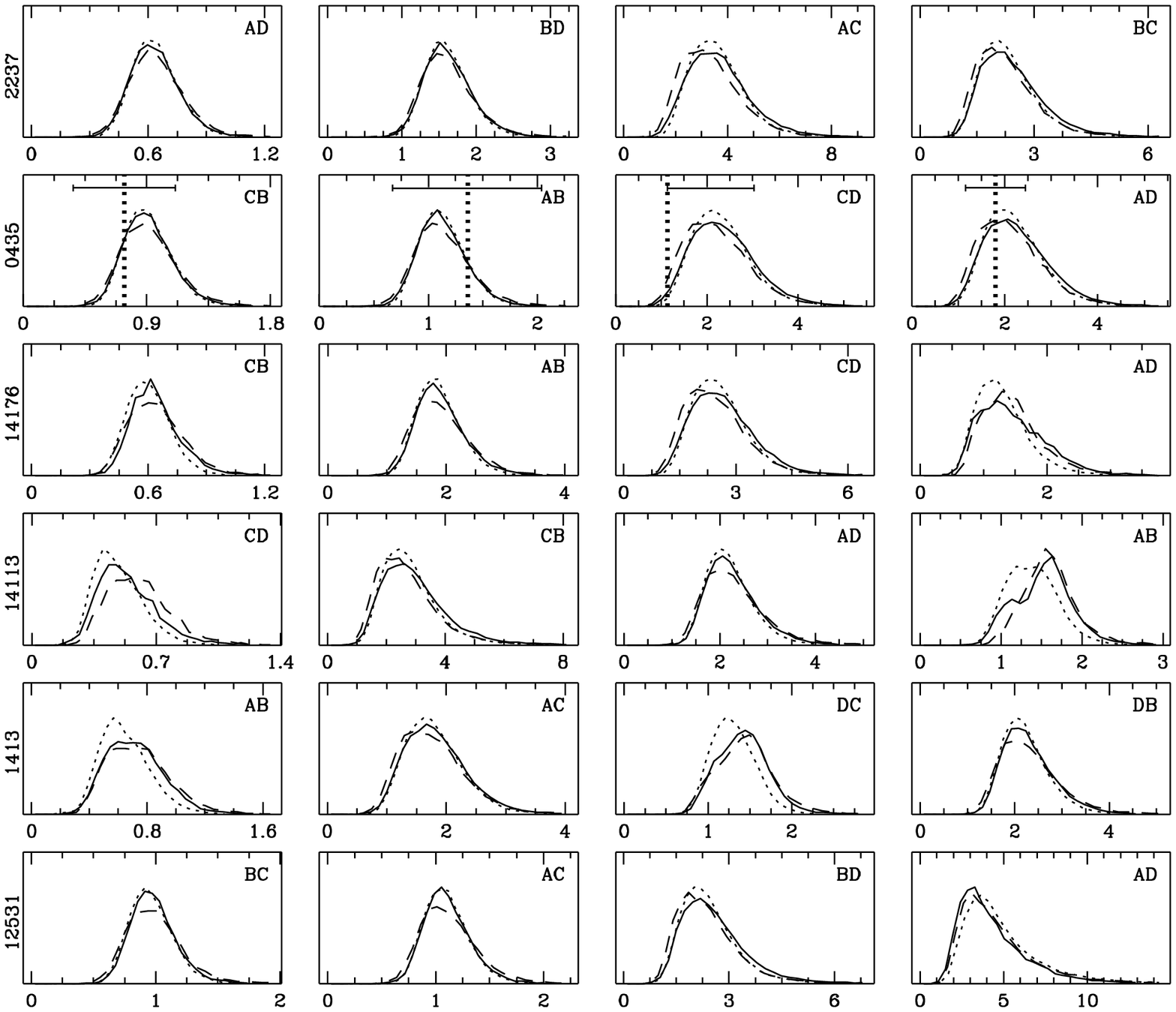}
\caption{
Same as Figure \ref{fig:trat-fold}, but for the known cross lenses.
\label{fig:trat-cross}}
\end{center}
\end{figure*}

To get a sense of the range of time delays that can be produced by smooth lens models, we show histograms of the scaled time delay (Figs.\ \ref{fig:tdel-fold}-\ref{fig:tdel-cross}) and time delay ratio (Figs.\ \ref{fig:trat-fold}-\ref{fig:trat-cross}) for each image pair in the 25 known four-image lenses. There are many lenses whose time delays have not yet been measured, but the histograms for those cases are still pedagogically useful and provide a way to predict what the time delay should be if the lens in question is not anomalous. In this subsection we discuss the general features of the histograms. In the following subsections we analyze the lenses with known time delays and make predictions for the remainder.

We divide the observed lenses into three groups: folds (Figs.\ \ref{fig:tdel-fold}, \ref{fig:tdel-fold2}, \ref{fig:trat-fold}, and \ref{fig:trat-fold2}), cusps (Figs.\ \ref{fig:tdel-cusp} and \ref{fig:trat-cusp}), and crosses (Figs.\ \ref{fig:tdel-cross} and \ref{fig:trat-cross}). We assign one of these canonical lens morphologies to each observed four-image lens according to the classification scheme of \citet{Keeton-cusp, Keeton-fold}. For each lens, we define $d_1^{\,*}$ to be the smallest value of the pairwise image separations. The lenses in each figure are arranged in rows such that $d_1^{\,*}$ increases from top to bottom. Within each row, the panels are arranged such that $d_1$ increases from left to right. As a check for systematic effects, we plot histograms of the time delays and time delay ratios produced by Monte Carlo simulations using galaxy samples from \citet{Bender-ellip}, \citet{Jorgensen-ellip}, and \citet{Saglia-ellip} (see \S\ref{sec:distributions}).

We first consider histograms of the scaled time delays, which are shown in Figures \ref{fig:tdel-fold}-\ref{fig:tdel-cross}.  A negative time delay indicates that the middle image in the triplet defined by $d_1$ and $d_2$ has negative parity (eq. [\ref{eq:minsadmin}]), while a positive time delay indicates a middle image with positive parity (eq. [\ref{eq:sadminsad}]).  The histograms are not highly sensitive to the input galaxy sample, especially in the tails (which are most important for identifying anomalies).  The histograms are somewhat asymmetric with a longer tail on the side away from zero, which presumably reflects the fact that the histograms are bounded by zero by construction.  Many of the histograms have a large width relative to the mean/median, which is not too surprising because we have been quite generous in matching mock lenses to observed lenses on the basis of minimal information (just $d_1$, $d_2$, and parity).  This is consistent with our goal of being conservative in identifying time delay anomalies.

Next, we consider histograms of the time delay ratios, which are shown in Figures \ref{fig:trat-fold}--\ref{fig:trat-cross}. These histograms are skewed to the right, presumably because the time delay ratios are positive and hence the histograms are bounded by zero on the left but unbounded on the right. The overall structure of the time delay ratio histograms does not vary much from one lens to another, or between image pairs of a given lens. This suggests that conclusions drawn from time delay ratios are not terribly sensitive to the lens morphology (fold, cusp, or cross), which may prove quite useful.

\subsection{Identifying Time Delay Anomalies in Observed Lenses}
\label{anomaly-id}

\begin{deluxetable*}{lccrcrc}
\tablewidth{0pt} \tabletypesize{\small} \tablecaption{P-values for
scaled time delays \label{tab:pval-scaled}} \tablehead{
Lens & Image & Rank & Obs. $\Delta t_1$ & Error    & P-value & P-values for\\
Name & Pair  &      & (days)            & Interval & for $\Delta
t_1$ & Err. Interval } \startdata
 1004 & BA & 1 & 40.6 & (35.2, 46.0) & 0.114 & (0.0630, 0.170) \\
 1004 & CA & 3 & -822. & (-828., -815.) & 0.917 & (0.914, 0.919) \\
 1115 & A1A2 & 1 & 0.149 & (0.131, 0.167) & 0.109 & (0.0670, 0.154) \\
 1115 & A1B & 2 & 11.7 & (8.10, 15.3) & 0.292 & (0.121, 0.487) \\
 1115 & CB & 3 & -25.0 & (-29.8, -20.2) & 0.983 & (0.922, 0.996) \\
 1115 & CA2 & 4 & -13.3 & (-16.3, -10.3) & 0.991 & (0.971, 0.998) \\
 1608 & AC & 1 & -4.50 & (-9.00, 0) & 1.00 & (0.995, 1.00) \\
 1608 & BC & 2 & -36.0 & (-40.5, -31.5) & 0.921 & (0.849, 0.965) \\
 1608 & AD & 3 & 45.5 & (41.0, 50.0) & 0 & (0, 0) \\
 1608 & BD & 4 & 77.0 & (71.0, 83.0) & --- & (---, ---) \\
 2033 & A1C & 2 & 27.1 & (14.2, 40.0) & 0.469 & (0.0882, 0.791) \\
 2033 & BA2 & 3 & -35.5 & (-39.7, -31.3) & 0.903 & (0.843, 0.945) \\
 2033 & BC & 4 & -62.6 & (-74.9, -50.3) & 0.998 & (0.995, 1.00) \\
 0911 & BA & 1 & 6.00 & (-24.0, 36.0) & 1.00 & (0, 1.00) \\
 0911 & BC & 2 & 5.00 & (-48.7, 58.7) & 1.00 & (0, 1.00) \\
 0911 & DC & 3 & -154. & (-202., -106.) & 1.00 & (1.00, 1.00) \\
 0911 & DA & 4 & -143. & (-161., -125.) & 1.00 & (1.00, 1.00) \\
 1131 & BA & 1 & -12.0 & (-16.5, -7.50) & 0 & (0, 0) \\
 1131 & CA & 2 & -9.60 & (-15.6, -3.60) & 0 & (0, 0.00601) \\
 1131 & BD & 3 & 99.0 & (75.0, 123.) & 0.455 & (0.266, 0.632) \\
 1131 & CD & 4 & 96.6 & (72.6, 121.) & 0.376 & (0.206, 0.548) \\
 1422 & AB & 1 & -1.50 & (-5.70, 2.70) & 0 & (0, 1.00) \\
 1422 & CB & 2 & -8.20 & (-14.2, -2.20) & 0 & (0, 0.000361) \\
 0435 & CB & 1 & -5.90 & (-8.30, -3.50) & 0.931 & (0.773, 0.991) \\
 0435 & AB & 2 & -8.00 & (-10.4, -5.60) & 0.862 & (0.676, 0.963) \\
 0435 & CD & 3 & 12.3 & (9.90, 14.7) & 0.0489 & (0.0205, 0.0907) \\
 0435 & AD & 4 & 14.4 & (11.7, 17.1) & 0.0668 & (0.0307, 0.119)
\enddata
\tablecomments{
Column 1 gives abridged lens names (see Table \ref{tab:lens-info} for the corresponding full names). Image pairs contain one minimum and one saddle. The labels in column 2 list the minimum image first. We rank image pairs according to their separation in column 3, with smaller numbers corresponding to smaller separations. Columns 4 and 5 list the observed time delays, along with their 3$\sigma$ error bars. In cases where the measurement uncertainties are asymmetric about the observed value, we create symmetric error bars with uncertainty $\sigma$, where $\sigma$ refers to the larger of the two measurement uncertainties. Note that the parity cut described in \S\ref{sec:distributions} specifies the sign of the time delay in a given image pair, so time delays with the opposite sign are unphysical even if such values are formally allowed by the measurement uncertainties. The observational time delay data used here can be found in Table 1 of \citet{Oguri_tdel}, except for the lenses 1004 \citep{Fohlmeister-1004-tdel-CA} and 2033 \citep{Vuissoz-2033-tdel}. Columns 6 and 7 give P-values for the time delays shown in columns 4 and 5, using the galaxy sample of \citet{Bender-ellip}.
}
\end{deluxetable*}

\begin{deluxetable*}{lccrcrc}
\tablewidth{0pt} \tabletypesize{\small} \tablecaption{P-values for
time delay ratios \label{tab:pval-rat}} \tablehead{
Lens & Image & Rank & Observed                & Error    & P-value & P-values for\\
Name & Pair  &      & $\Delta t_1/\Delta t_2$ & Interval & for
$\Delta t_1/\Delta t_2$ & Err. Interval } \startdata
 1004 & CA & 3 & 20.2 & (17.5, 22.9) & 0.0646 & (0.0139, 0.161) \\
 1115 & A1A2 & 1 & 0.0127 & (0.00853, 0.0169) & 0.178 & (0.0253, 0.445) \\
 1115 & A1B & 2 & 78.5 & (52.6, 104.) & 0.751 & (0.364, 0.916) \\
 1115 & CB & 3 & 2.14 & (1.36, 2.91) & 0.755 & (0.0543, 0.962) \\
 1115 & CA2 & 4 & 89.3 & (66.4, 112.) & 0.344 & (0.130, 0.527) \\
 1608 & AC & 1 & 0.125 & (-0.000973, 0.251) & 0 & (0, 0.696) \\
 1608 & BC & 2 & 8.00 & (-0.0623, 16.1) & 0.706 & (0, 0.987) \\
 1608 & AD & 3 & 10.1 & (-0.0493, 20.3) & 0.177 & (0, 0.967) \\
 1608 & BD & 4 & 2.14 & (1.82, 2.45) & --- & (---, ---) \\
 2033 & BC & 4 & 2.31 & (1.12, 3.50) & 0.193 & (0, 0.746) \\
 0911 & BA & 1 & 1.20 & (-13.0, 15.4) & 0.992 & (0, 1.00) \\
 0911 & BC & 2 & 0.833 & (-9.04, 10.7) & 0.00753 & (0, 1.00) \\
 0911 & DC & 3 & 30.8 & (-300., 362.) & 0 & (0, 0.987) \\
 0911 & DA & 4 & 23.8 & (-95.4, 143.) & 0 & (0, 0.0200) \\
 1131 & BA & 1 & 1.25 & (0.339, 2.16) & 0.976 & (0, 1.00) \\
 1131 & CA & 2 & 0.800 & (0.217, 1.38) & 0.0240 & (0, 0.785) \\
 1131 & BD & 3 & 8.25 & (4.57, 11.9) & 0.000404 & (0, 0.00623) \\
 1131 & CD & 4 & 10.1 & (3.29, 16.8) & 0.00388 & (0, 0.0589) \\
 1422 & AB & 1 & 0.183 & (-0.346, 0.712) & 0.0350 & (0, 1.00) \\
 1422 & CB & 2 & 5.47 & (-10.4, 21.3) & 0.965 & (0, 1.00) \\
 0435 & CB & 1 & 0.738 & (0.365, 1.11) & 0.157 & (0.000215, 0.861) \\
 0435 & AB & 2 & 1.36 & (0.671, 2.04) & 0.843 & (0.00951, 0.996) \\
 0435 & CD & 3 & 2.08 & (1.14, 3.03) & 0.378 & (0.0151, 0.826) \\
 0435 & AD & 4 & 1.80 & (1.16, 2.44) & 0.275 & (0.0216, 0.635)
\enddata
\tablecomments{
Columns 1--3 have the same meaning as in Table \ref{tab:pval-scaled}. Columns 4 and 5 list time delay ratios and their corresponding error intervals. Errors on time delay ratios are computed by propagating errors from observed time delays (see \S\ref{anomaly-id} for details). Note that negative values in column 5 are unphysical and have only formal meaning (see Table \ref{tab:pval-scaled}). Columns 6 and 7 give P-values for the time delay ratios shown in columns 4 and 5, using the galaxy sample of \citet{Bender-ellip}.
}
\end{deluxetable*}

Eight of the 25 known four-image lenses have at least one image pair with an observed time delay (see Tables \ref{tab:pval-scaled} and \ref{tab:pval-rat}). We can use our simulations to determine whether these lenses are consistent with lensing by an ellipsoidal mass distribution with tidal shear. Specifically, we compare the observed value of the time delay and (if available)\footnote{Recall that to construct the time delay ratio we must know not only the time delay for the image pair, but also the time delay to the next nearest image.} the time delay ratio with our predicted distributions.  If the observed value lies outside of the predicted range, we classify the time delay as anomalous and interpret it as strong evidence that the lens galaxy cannot be described as a simple, relatively smooth mass distribution with shear.

This statement is quantified by the statistical P-value, which gives the fraction of matching mock lenses whose time delays (or ratios) are smaller than the observed value; either $P>0.995$ or $P<0.005$, for example, would indicate that a time delay is anomalous at more than 99\% confidence. We consider $P>0.995$ or $P<0.005$ to indicate a strong anomaly, and $0.975<P<0.995$ or $0.005<P<0.025$ to indicate a marginal anomaly.  Given the relatively small number of known four-image lenses, we should not read too deeply into marginal anomalies since a few outliers are only to be expected.  The extreme P-values required of strong anomalies, however, would be difficult to interpret as statistical flukes, so conclusions based on such systems should be robust.  The choice of input galaxy sample for the Monte Carlo simulations has no significant effect on our conclusions; so for simplicity we report P-values computed using only the sample of \citet{Bender-ellip}, which is the larger of the two samples that include $a_4$ measurements.

To assess whether measurement errors in observed time delays affect our results, we compute P-values for the endpoints of the observed $3\sigma$ error interval.  This is a simple task for time delays, but for time delay ratios we first need to propagate errors from the time delays into the ratios. For time delay errors that are symmetric, we assume they are Gaussian and propagate them using the standard formula for a quotient.  If the errorbars are asymmetric, we conservatively take the larger errorbar as the Gaussian standard deviation, $\sigma$. While this approach is not strictly correct, it is the best we can do without knowing the full error distribution, and it is conservative in the sense that it should overestimate the uncertainties in the time delay ratio. It turns out that our identifications of anomalies are not affected by the observational errors in most cases (see Tables \ref{tab:pval-scaled} and \ref{tab:pval-rat}, and the following discussions of individual lenses). This does not mean, however, that there is no need to measure time delays more precisely. Rather, it reflects the fact that our predicted time delay distributions are quite broad because we have deliberately chosen to be generous in matching observed and mock lenses. When it comes to analyzing time delay anomalies to extract physical information (about the substructure mass function, for example), it will be necessary to do detailed modeling for which precise image positions and time delays will be vital (see \citealt{Keeton-tdel} and \citealt{Moustakas-astro2010} for more discussion).

We now discuss all of the lenses with at least one observed time delay in the order in which they appear in Figures \ref{fig:tdel-fold}--\ref{fig:tdel-cross}.
\newline
\newline
{\it PG 1115+080.}
All of the differential time delays in the fold lens 1115 are known \citep{Schechter-1115,Barkana-1115,Chartas-1115}. The flux ratio between the close images A1 and A2 is anomalous at optical and X-ray wavelengths \citep[see][and references therein]{Pooley-fluxrat}. The mid-IR flux ratios are not anomalous, however, which suggests that the X-ray/optical anomaly is caused by microlensing \citep{Chiba-IR}. Since microlensing does not affect time delays \citep{Keeton-tdel}, we might expect the time delays not to be anomalous. Indeed, Figure \ref{fig:tdel-fold} and Tables \ref{tab:pval-scaled} and \ref{tab:pval-rat} show that the A1A2 time delay is not anomalous.  We now turn to the time delays between the ``distant'' image pairs.  The P-values of the pairs CA2 and CB would be anomalous if we considered only the $1\sigma$ errorbars, but they fall below the threshold when we use $3\sigma$ errorbars.  Since there are no time-delay anomalies in 1115, we conclude that the observed optical flux anomalies are indeed due to microlensing.
\newline
\newline
{\it SDSS J1004+4112.}
The fold lens 1004 is produced by a cluster of galaxies and contains five lensed images whose temporal ordering is C-B-A-D-E \citep{Fohlmeister-1004-tdel-CA}. Images C and B are minima, A and D are saddles, and E is a maximum \citep{Inada-1004-central}. This is the only quad lens whose maximum (doubly-negative parity) image has been observed, so we have not included maxima in our analysis. The time delays are known for the image pairs BA and CA, so we can examine the corresponding scaled time delays and also the time delay ratio for the pair CA; but the other time delays and ratios have not yet been determined. The P-values for the known time delays do not indicate anomalies (according to our 95\% confidence criterion). This seems surprising, because the actual lens potential is presumably very different from our assumed model of an isothermal galaxy with shear: clusters are not expected to have isothermal profiles \citep[e.g.,][]{NFW-97}, and the galaxies in the cluster create significant complexity in the potential.  We believe the results for 1004 indicate that our method for finding time delay anomalies is conservative.
\newline
\newline
{\it WFI 2033$-$4723.}
For the fold lens 2033, \citet{Vuissoz-2033-tdel} report the time delays between images B and C and between B and the combination of the close images A1 and A2.  The time delay between A1 and A2 was too small to be measured; it is expected to be short enough that we can assume A1 and A2 have effectively the same light travel time for the purpose of determining the time delays for the image pairs A1C and BA2. Among the distant image pairs, the BC pair has the largest separation and also a strong time delay anomaly, while the two pairs with smaller distances are not anomalous. The nominal value of the time delay ratio for the BC pair does not indicate an anomaly, although the uncertainties are large enough that the situation is not conclusive at present.

The time delay in BC may be affected by the group of galaxies of which the main lens is a member \citep{2026-2033-discovery,Vuissoz-2033-tdel}.  We find two lines of evidence supporting this hypothesis.  First, the time delays become more anomalous as the image separation increases.  Second, our predicted time delays are longer than the observed time delays; adding environmental effects to the lens model would generally reduce the predicted time delays.  A group would contribute a non-negative convergence $\kappa_{\rm env}$ to the lens potential, which would rescale the predicted time delays by a factor $1-\kappa_{\rm env} < 1$, and it could also create higher-order effects that may be more complicated \citep{Keeton-env}.  This and alternative hypotheses that the time delays are affected by a change in the radial profile of the lens galaxy or even the global value of the Hubble constant are discussed below.
\newline
\newline
{\it B1608+656.}
All of the differential time delays are known for the fold lens 1608 \citep{Fassnacht-1608-tdel}. Because there are two galaxies inside the Einstein angle, this lens is not necessarily expected to be well described by our models. Indeed, it is extremely hard for single-galaxy models (even with shear) to reproduce the $d_1$ and $d_2$ values for the image pair BD: we find only one mock image pair that matches the observed values. We are therefore unable to compute P-values for the BD time delay (cf.\ Table \ref{tab:pval-scaled}).

Among the other image pairs, the fold pair AC and the distant pair AD are anomalous in terms of their scaled time delays.  All of these anomalies are very strong; in fact, the AD scaled time delay and its $3\sigma$ values have P-values that are strictly zero, meaning there are no matching mock lenses whose time delays are more extreme than the observed values. We note that these anomalies do not necessarily reveal CDM substructure, because the presence of two lens galaxies makes the lens potential more complicated than we have allowed for here. Inferences about substructure need to be done in the context of models that treat this complex lens in more detail \citep[e.g.,][]{Koopmans-1608-model,Suyu-1608-model}.
\newline
\newline
{\it RX J0911+0551.}
We now turn to the cusp lenses. \citet{Hjorth-0911-tdel} report the time delays between each of the three cusp images (A, B, and C) and the fourth image (D) for 0911. We find clear evidence of time delay anomalies in the DC and DA image pairs, which are the two pairs with the largest separations. Since the lens galaxy in 0911 is part of a cluster \citep{0911-cluster}, it is possible that these anomalies are due to environmental effects.  Pinning down the origin of the anomalies (i.e., the environment of the lens, or substructure) will require that the time delays among the cusp images be precisely measured.\footnote{\citet{Chartas-0911-xray} and \citet{Morgan_1131} have measured time delays among close images in other lens systems, so we are hopeful that it will be possible to do so in 0911.}  Although the close time delays can be inferred from current data, the P-values they predict span the range (0, 1) when 3$\sigma$ measurement uncertainties are included.  At this point we conclude that the 0911 time delays are very intriguing and warrant further study.
\newline
\newline
{\it RX J1131$-$1231.}
\citet{Morgan_1131} report the time delays among the close images A, B, and C for the cusp lens 1131, along with an estimate of the time delay to the distant image D. They note that the time delays among the close images are much longer than expected for a smooth mass distribution, and suggest that the time delays indicate the presence of a massive $\left(\sim 5 \times 10^{10} M_\odot\right)$ clump near image A. \citet{Keeton-tdel} highlight a second peculiarity of the 1131 time delays, namely that the minimum image B is observed to lead the minimum image C, whereas smooth models predict the reverse. They suggest that a population of clumps could reverse the temporal ordering of the two minimum images in a cusp lens.

We cannot directly address the issue of the temporal ordering of images B and C, because we do not consider the time delay between images with the same parity. However, we can offer a more model-independent assessment of the time delays for the other image pairs. We find that the time delays for the cusp pairs BA and CA are anomalous: accounting for 3$\sigma$ errorbars, the P-value for BA is strictly zero, while the pair CA falls just short of a strong anomaly (0.006 compared with the threshold of 0.005). This strengthens the conclusion by \citet{Morgan_1131} that the observed time delays are not at all consistent with lensing by a reasonable smooth mass distribution.  Interestingly, we also find that the distant image pair BD has an (almost strongly) anomalous time-delay ratio, even though its scaled time delay is not anomalous. We note that this anomaly is caused by the observed value being smaller than most of the predicted values; this might be understood in terms of the BA time delay, which appears in the denominator of the BD time-delay ratio, being anomalously longer (in absolute value) than expected. Our main conclusion is that the 1131 time delays are highly anomalous, and this general conclusion together with the specific analyses of \citet{Morgan_1131} and \citet{Keeton-tdel} strongly suggests that this lens contains significant substructure.  We also note, however, that we must be careful when interpreting time-delay ratios, since an anomaly may result from the denominator being anomalous rather than the pair of interest.
\newline
\newline
{\it B1422+231.}
For 1422, the nominal values of the time delays \citep{Patnaik-1422-tdel} among the three cusp images A, B, and C indicate strong anomalies. However, the uncertainties in the claimed time delays are not much smaller than the time delays themselves, so conclusive statements are impossible at this point. (Notice, for example, that the range of P-values for the AB pair spans 0 to 1.00 given the uncertainties.)  The system clearly warrants further study, especially since it is well-known to have anomalous flux ratios \citep{Mao-flux, Bradac-1422}.
\newline
\newline
{\it HE 0435$-$1223.}
Finally, we consider the cross lens 0435, whose time delays were measured by \citet{Kochanek-0435-rising}. The cross image configuration features no close pairs of images, so all the time delays could be affected by substructure or large-scale complexity in the lens potential, or both.  We do not find anomalies in any of the image pairs, suggesting that the anomalous fluxes at the $\sim$0.2 mag level in images A and C are most likely due to microlensing \citep{Kochanek-0435-rising} rather than substructure \citep{Morgan-0435}.  Given our goal of being conservative, it is reassuring to find that a cross lens like 0435 has perfectly reasonable time delays.
\newline
\newline
In the preceding discussion of individual lenses we have interpreted time delay anomalies as evidence for complex structure in the lens potential. Here we briefly consider three alternative interpretations.  One possibility is that the Hubble constant, which is needed to compare observed and predicted time delays, differs from our assumed value of $H_0 = 70$ km s$^{-1}$ Mpc$^{-1}$.  A second possibility is that a lens has a mass sheet, or external convergence $\kappa_{\rm ext}$, whose effects we have neglected \citep{Keeton-env}.  The apparent anomalies in 2033 and 0911 arise because the model predictions are longer than the observed time delays. Since predicted time delays scale as $\Delta t \propto (1-\kappa_{\rm ext}) H_0^{-1}$, we would need either to have a strong \emph{negative} mass sheet or to increase $H_0$ in order to eliminate the anomalies.  Focusing on the $H_0$ possibility, in order to reduce the predicted time delays such that the 95\% confidence intervals overlap the observed values in 2033 and 0911, we would require an $H_0$ value in excess of 100 km s$^{-1}$ Mpc$^{-1}$, which does not seem like a viable option. The lens 1131 is a bit different, because the predicted time delays are shorter than the observed values, which would imply a positive mass sheet or a smaller Hubble constant than we have assumed.  Regardless of whether the required mass sheet or $H_0$ value is reasonable, however, the existing evidence strongly implies that 1131 does contain substructure \citep{Morgan_1131,Keeton-tdel}.

A third possibility is that the radial density profiles of the lens galaxies differ from the isothermal profile we have assumed. Changing the radial profile has more complicated effects on the predicted time delays, but \citet{Kochanek-tdel} argues that to leading order the changes can be approximated with the scaling $\Delta t \propto (1-\langle\kappa\rangle)$, where $\langle\kappa\rangle$ is the mean convergence in the vicinity of the Einstein angle (specifically, in the annulus spanned by the images).  Using this scaling, we can estimate the radial profile that would be required if we wanted to reduce the predicted time delays so the 95\% confidence intervals overlap the observed values.  (This analysis only yields an estimate, because the scaling is only approximate and does not include the full complexity of ellipsoidal mass distributions; but it is still instructive.)  If we write the 3-d density profile as $\rho \propto r^{-\eta}$, we estimate that we would need $\eta \approx 1.7$ for 2033, and $\eta \approx 1.5$ for 0911, compared to $\eta=2$ for an isothermal distribution.  For comparison, \citet{Koopmans-SLACS-III} find that lens galaxies have a mean power law index of $\langle\eta\rangle = 2.01_{-0.03}^{+0.02}$, with an RMS scatter of 0.12.  Again, the simple analysis here needs to be interpreted with some care, but it does suggest that changing the radial profile does not provide a compelling explanation for the apparent time delay anomalies.

\subsection{Predictions for the Remaining Lenses}

To complete our analysis, we present predictions of the scaled time delays (Table \ref{tab:tdel-scaled}) and time delay ratios (Table \ref{tab:tdel-rat}) for all mixed-parity image pairs in all 25 known four-image lenses. Specifically, we use the simulations based on the galaxy sample of \citet{Bender-ellip} to compute the median value, 95\% confidence interval, and 99\% confidence interval for each quantity for each image pair. These results give a sense of what the time delays should be for lenses that are adequately described as ellipsoidal mass distributions with tidal shear. (See \citealt{Saha-tdel-predict} for a complementary approach based on fitting pixellated mass models to individual lenses.) There is currently great interest in lens monitoring \citep[e.g.,][]{Courbin-tdel,COSMOGRAIL-1,Kochanek-SCMA,OMEGA,Moustakas-astro2010}, and we hope our predictions will be useful in planning observational campaigns to measure time delays. Furthermore, as new time delays are measured, it will be a simple matter to compare them with our predictions to determine whether the time delays are anomalous in a way that indicates a complex lens potential (due either to substructure or to the lens environment).

\section{CONCLUSIONS}
\label{sec:conclusions}

We have introduced a new method to use gravitational lens time delays to detect complex structure in the lens potential.  The complexity may be associated with CDM substructure, in which case time delays offer the chance to learn more about the substructure population than is possible with lens flux ratios; or it may be associated with the lens environment, such as a group or cluster of galaxies surrounding the lens.  The basic approach is to determine the range of time delays that can be produced by reasonable smooth lens models, so that we can identify outliers as being anomalous. To get a sense of how this could work, we first studied the dependence of the time delay between the close pair of images in a fold lens on the position of the source and the form of the lens potential. For a source near a fold point, we have found that the time delay remains approximately constant as the source moves along the caustic. For a lens modeled by an elliptical galaxy with $m=4$ multipole perturbations and tidal shear, the time delay increases with ellipticity and shear, but is not very sensitive to $m=4$ modes.

Using Monte Carlo simulations, we then constructed distributions of the time delays in four-image lenses. This approach can handle fold, cusp, and cross lenses, which comprise the three canonical four-image lens morphologies. By constructing a catalog of mock lenses based on observed populations of elliptical galaxies, we computed the range of time delays and time delay ratios that would be expected for a smooth lens potential (i.e., one with ellipticity, shear, and $m=4$ multipoles). By comparing observed time delays with the predicted ranges, we have found time delay anomalies in the systems RX J0911+0551, RX J1131$-$1231, B1422+231, B1608+656, and WFI 2033$-$4723.  It is unlikely that these anomalies can be explained by errors in our assumed values of the Hubble constant or the slope of the density profile: the Hubble constant would have to be unreasonably high or the density profile surprisingly shallow in order to explain some of the apparent anomalies (we specifically discussed RX J0911+0551 and WFI 2033$-$4723), and neither possibility could explain RX J1131$-$1231.  It is possible to further reduce sensitivity to the Hubble constant and density profile by working with time delay ratios, although in general we have found that time delay ratios have less power to reveal anomalies.  Part of the problem is that there are fewer time delay ratios known than time delays themselves, and a large uncertainty in a given time delay will lead to a correspondingly large uncertainty in the ratio, making definitive conclusions difficult.

In general, anomalies between close pairs of images in fold and cusp lenses should provide the cleanest evidence of substructure. The cusp lens RX J1131$-$1231 contains such anomalies, which is consistent with conclusions based on more detailed modeling by \citet{Morgan_1131} and \citet{Keeton-tdel}. The cusp lenses B1422+231 and RX J0911+0551 show evidence of time delay anomalies in the cusp triplet, but the large uncertainties in the measured time delays prevent firm conclusions at present. The only fold pair that is clearly anomalous is in B1608+656, but the peculiar nature of this system (with two lens galaxies inside the Einstein angle) makes it difficult to draw a definitive conclusion about substructure from our analysis.  Among larger-separation image pairs we have found time delay anomalies in the fold lens WFI 2033$-$4723 and the cusp lens RX J0911+0551.  Both lenses show evidence of a complex environment, but in order to distinguish between that and substructure as the origin of the time delay anomalies it will be necessary to precisely measure the time delays between the close images (the fold pair in WFI 2033$-$4723, and the cusp triplet in RX J0911+0551).

In the hope that the sample of observed precision time delays will continue to grow, we have predicted the time delays for all mixed-parity image pairs in all 25 known four-image lenses.  As new time delays are measured, it will be a simple matter to compare them with our predicted confidence intervals to determine whether they are anomalous.  If a lens galaxy contains complex structure, time delays should help reveal it.  Flux ratios provide a powerful way to find small-scale structure, but they are not unique in this; time delays hold great promise for contributing to our understanding of the role played by dark matter in the universe, especially when combined with other lensing observables.

\acknowledgements
We thank Leonidas Moustakas and Ross Fadely for helpful conversations.  We also thank the referee, Olaf Wucknitz, for several helpful suggestions that improved the manuscript.  ABC would also like to thank Mark Eichenlaub and Dennis Lam for their input.
Part of this work was funded by NSF grant AST-0747311.
ABC is currently supported by an appointment to the NASA Postdoctoral Program at the Jet Propulsion Laboratory, administered by Oak Ridge Associated Universities through a contract with NASA.


\newpage


\newpage

\newpage

\newpage

\newpage

\newpage

\newpage

\newpage

\newpage


\LongTables
\newpage

\begin{deluxetable*}{lllrcc}
\tablewidth{0pt} \tabletypesize{\small} \tablecaption{Median values
and confidence intervals for scaled time delays \label{tab:tdel-scaled}}
\tablehead{
Lens & Image & $d_1$    & Median       & 95\% Conf. & 99\% Conf. \\
Name & Pair  & (arcsec) & $\Delta t_1$ & Interval   & Interval }
\startdata
 0128 & {AB[D]} & 0.14 & 0.0164 & {(0.00598,0.0403)} &
{(0.00386,0.0483)} \\
 0128 & {AD[B]} & 0.27 & 0.152 & {(0.0527,0.332)} &
{(0.0333,0.394)} \\
 0128 & {CD[A]} & 0.42 & -1.02 & {(-1.22,-0.800)} &
{(-1.26,-0.716)} \\
 0128 & {CB[A]} & 0.5 & -1.43 & {(-1.72,-1.14)} & {(-1.78,-1.03)}
\\
 0230 & {AB[D]} & 0.74 & 0.0745 & {(0.0583,0.0954)} &
{(0.0547,0.104)} \\
 0230 & {CD[A]} & 1.46 & -0.654 & {(-0.806,-0.537)} &
{(-0.848,-0.507)} \\
 0230 & {AD[B]} & 1.64 & 1.56 & {(0.728,1.99)} & {(0.436,2.05)}
\\
 0230 & {CB[A]} & 1.65 & -0.583 & {(-0.792,-0.349)} &
{(-0.858,-0.271)} \\
 0414 & {A1A2[B]} & 0.41 & -0.00223 & {(-0.00471,-0.000827)} &
{(-0.00553,-0.000520)} \\
 0414 & {BA2[A1]} & 1.71 & -0.129 & {(-0.212,-0.0560)} &
{(-0.248,-0.0383)} \\
 0414 & {A1C[A2]} & 1.96 & 0.635 & {(0.178,1.50)} & {(0.112,1.66)}
\\
 0414 & {BC[A2]} & 2.13 & 1.10 & {(0.668,1.69)} & {(0.536,1.97)}
\\
 0712 & {AB[C]} & 0.17 & -0.000571 & {(-0.00136,-0.000185)} &
{(-0.00161,-0.000148)} \\
 0712 & {CB[A]} & 0.91 & -0.0627 & {(-0.101,-0.0272)} &
{(-0.113,-0.0203)} \\
 0712 & {CD[B]} & 1.18 & 0.484 & {(0.177,0.911)} & {(0.118,1.12)}
\\
 0712 & {AD[B]} & 1.25 & 0.697 & {(0.177,1.66)} & {(0.105,1.77)}
\\
 0810 & {AB[C]} & 0.18 & -0.00195 & {(-0.00407,-0.000712)} &
{(-0.00474,-0.000503)} \\
 0810 & {CB[A]} & 0.69 & -0.0868 & {(-0.144,-0.0377)} &
{(-0.160,-0.0274)} \\
 0810 & {AD[B]} & 0.84 & 0.494 & {(0.147,1.19)} & {(0.0950,1.46)}
\\
 0810 & {CD[B]} & 0.85 & 0.498 & {(0.188,0.920)} & {(0.128,1.13)}
\\
 0924 & {AD[C]} & 0.69 & 0.0297 & {(0.0108,0.0688)} &
{(0.00737,0.0816)} \\
 0924 & {AC[D]} & 1.18 & 0.141 & {(0.0494,0.303)} &
{(0.0332,0.353)} \\
 0924 & {BD[A]} & 1.46 & -0.170 & {(-0.298,-0.0738)} &
{(-0.334,-0.0506)} \\
 0924 & {BC[A]} & 1.53 & -0.370 & {(-0.549,-0.160)} &
{(-0.597,-0.108)} \\
 1004 & {BA[D]} & 3.73 & 0.0104 & {(0.00424,0.0225)} &
{(0.00307,0.0254)} \\
 1004 & {BD[A]} & 11.44 & 0.367 & {(0.115,0.839)} &
{(0.0748,1.10)} \\
 1004 & {CA[B]} & 11.84 & -0.203 & {(-0.344,-0.0912)} &
{(-0.383,-0.0629)} \\
 1004 & {CD[B]} & 14.38 & -1.04 & {(-1.40,-0.797)} &
{(-1.55,-0.698)} \\
 1115 & {A1A2[B]} & 0.48 & 0.00559 & {(0.00195,0.0147)} &
{(0.00136,0.0170)} \\
 1115 & {A1B[A2]} & 1.67 & 0.316 & {(0.0995,0.730)} &
{(0.0657,0.976)} \\
 1115 & {CB[A1]} & 1.99 & -0.783 & {(-1.04,-0.529)} &
{(-1.19,-0.437)} \\
 1115 & {CA2[A1]} & 2.16 & -0.525 & {(-0.716,-0.325)} &
{(-0.794,-0.249)} \\
 1330 & {AB[C]} & 0.43 & -0.00355 & {(-0.00739,-0.00129)} &
{(-0.00867,-0.000872)} \\
 1330 & {CB[A]} & 1.53 & -0.135 & {(-0.222,-0.0595)} &
{(-0.259,-0.0419)} \\
 1330 & {CD[B]} & 1.64 & 0.454 & {(0.175,0.718)} &
{(0.117,0.823)} \\
 1330 & {AD[B]} & 1.65 & 0.428 & {(0.130,1.01)} & {(0.0815,1.26)}
\\
 1555 & {AB[C]} & 0.09 & -0.00188 & {(-0.00397,-0.000676)} &
{(-0.00451,-0.000449)} \\
 1555 & {CB[A]} & 0.35 & -0.0933 & {(-0.155,-0.0410)} &
{(-0.171,-0.0301)} \\
 1555 & {AD[B]} & 0.4 & 0.406 & {(0.124,0.975)} & {(0.0785,1.23)}
\\
 1555 & {CD[B]} & 0.42 & 0.525 & {(0.204,0.951)} & {(0.135,1.17)}
\\
 1608 & {AC[B]} & 0.87 & -0.142 & {(-0.184,-0.111)} &
{(-0.202,-0.106)} \\
 1608 & {BC[A]} & 1.51 & -0.598 & {(-0.789,-0.351)} &
{(-0.856,-0.267)} \\
 1608 & {AD[C]} & 1.69 & 1.91 & {(1.23,2.21)} & {(1.10,2.27)} \\
1608 & {BD[C]} & 2.00 & 1.19 & $-$ & $-$ \\
 1933 & {4\_3[6]} & 0.46 & 0.0407 & {(0.0144,0.0888)} &
{(0.00911,0.104)} \\
 1933 & {4\_6[3]} & 0.63 & 0.119 & {(0.0413,0.252)} &
{(0.0271,0.294)} \\
 1933 & {1\_3[4]} & 0.9 & -0.364 & {(-0.559,-0.171)} &
{(-0.618,-0.120)} \\
 1933 & {1\_6[4]} & 0.91 & -0.488 & {(-0.672,-0.238)} &
{(-0.732,-0.167)} \\
 2026 & {A1A2[C]} & 0.33 & 0.00601 & {(0.00198,0.0154)} &
{(0.00137,0.0198)} \\
 2026 & {A1C[A2]} & 0.83 & 0.0929 & {(0.0308,0.218)} &
{(0.0200,0.258)} \\
 2026 & {BC[A1]} & 1.19 & -0.424 & {(-0.609,-0.195)} &
{(-0.664,-0.135)} \\
 2026 & {BA2[A1]} & 1.28 & -0.372 & {(-0.561,-0.188)} &
{(-0.624,-0.127)} \\
 2033 & {A1A2[C]} & 0.72 & 0.0155 & {(0.00572,0.0392)} &
{(0.00372,0.0469)} \\
 2033 & {A1C[A2]} & 1.54 & 0.182 & {(0.0623,0.392)} &
{(0.0421,0.463)} \\
 2033 & {BA2[A1]} & 2.01 & -0.348 & {(-0.537,-0.172)} &
{(-0.586,-0.116)} \\
 2033 & {BC[A1]} & 2.13 & -0.806 & {(-1.00,-0.579)} &
{(-1.07,-0.494)} \\
 0911 & {BA[C] - A2A1[A3]} & 0.48 & 0.00401 & {(0.00128,0.0104)} &
{(0.000726,0.0134)} \\
 0911 & {BC[A] - A2A3[A1]} & 0.62 & 0.00776 & {(0.00243,0.0208)} &
{(0.00178,0.0281)} \\
 0911 & {DC[B] - BA3[A2]} & 2.96 & -2.70 & {(-2.88,-2.38)} &
{(-2.90,-2.24)} \\
 0911 & {DA[B] - BA1[A2]} & 3.08 & -2.93 & {(-3.11,-2.56)} &
{(-3.12,-2.47)} \\
 1131 & {BA[C]} & 1.19 & -0.00680 & {(-0.0130,-0.00271)} &
{(-0.0150,-0.00185)} \\
 1131 & {CA[B]} & 1.26 & -0.00813 & {(-0.0148,-0.00331)} &
{(-0.0171,-0.00230)} \\
 1131 & {BD[A]} & 3.14 & 0.492 & {(0.152,1.09)} & {(0.0971,1.33)}
\\
 1131 & {CD[A]} & 3.18 & 0.533 & {(0.162,1.17)} & {(0.0991,1.40)}
\\
 1251 & {BA[C]} & 0.44 & 0.00435 & {(0.00139,0.0109)} &
{(0.000879,0.0131)} \\
 1251 & {BC[A]} & 0.7 & 0.0153 & {(0.00510,0.0388)} &
{(0.00339,0.0532)} \\
 1251 & {DA[B]} & 1.72 & -0.366 & {(-0.553,-0.182)} &
{(-0.619,-0.122)} \\
 1251 & {DC[B]} & 1.77 & -0.519 & {(-0.728,-0.291)} &
{(-0.802,-0.220)} \\
 1422 & {AB[C]} & 0.5 & -0.00981 & {(-0.0181,-0.00374)} &
{(-0.0204,-0.00261)} \\
 1422 & {CB[A]} & 0.82 & -0.0381 & {(-0.0655,-0.0153)} &
{(-0.0726,-0.0103)} \\
 1422 & {AD[B]} & 1.25 & 0.360 & {(0.118,0.808)} &
{(0.0767,1.07)} \\
 1422 & {CD[B]} & 1.29 & 0.413 & {(0.145,0.830)} &
{(0.0947,1.05)} \\
 2045 & {AB[C]} & 0.28 & -0.000236 & {(-0.000611,-0.0000811)} &
{(-0.000745,-0.0000684)} \\
 2045 & {CB[A]} & 0.56 & -0.00151 & {(-0.00294,-0.000679)} &
{(-0.00339,-0.000554)} \\
 2045 & {AD[B]} & 1.91 & 0.398 & {(0.112,1.00)} & {(0.0693,1.24)}
\\
 2045 & {CD[B]} & 1.93 & 0.443 & {(0.134,1.02)} & {(0.0804,1.26)}
\\
 0435 & {CB[A]} & 1.53 & -0.107 & {(-0.202,-0.0437)} &
{(-0.227,-0.0289)} \\
 0435 & {AB[C]} & 1.59 & -0.119 & {(-0.224,-0.0484)} &
{(-0.251,-0.0329)} \\
 0435 & {CD[B]} & 1.85 & 0.261 & {(0.0997,0.519)} &
{(0.0667,0.598)} \\
 0435 & {AD[B]} & 1.88 & 0.277 & {(0.106,0.544)} &
{(0.0718,0.623)} \\
 12531 & {BC[A]} & 0.77 & -0.156 & {(-0.269,-0.0696)} &
{(-0.297,-0.0466)} \\
 12531 & {AC[B]} & 0.78 & -0.160 & {(-0.283,-0.0688)} &
{(-0.318,-0.0460)} \\
 12531 & {BD[C]} & 0.91 & 0.330 & {(0.128,0.634)} &
{(0.0856,0.726)} \\
 12531 & {AD[C]} & 1.02 & 0.749 & {(0.297,1.30)} & {(0.201,1.58)}
\\
 14113 & {CD[B]} & 1.13 & 0.248 & {(0.169,0.341)} &
{(0.135,0.376)} \\
 14113 & {CB[D]} & 1.38 & 0.344 & {(0.129,0.661)} &
{(0.0869,0.762)} \\
 14113 & {AD[C]} & 1.41 & -0.311 & {(-0.486,-0.133)} &
{(-0.538,-0.0898)} \\
 14113 & {AB[C]} & 1.42 & -0.461 & {(-0.641,-0.192)} &
{(-0.729,-0.133)} \\
 1413 & {AB[C]} & 0.76 & 0.208 & {(0.118,0.330)} &
{(0.0870,0.361)} \\
 1413 & {AC[B]} & 0.87 & 0.239 & {(0.0941,0.478)} &
{(0.0620,0.547)} \\
 1413 & {DC[A]} & 0.91 & -0.354 & {(-0.473,-0.165)} &
{(-0.509,-0.107)} \\
 1413 & {DB[A]} & 0.96 & -0.320 & {(-0.498,-0.135)} &
{(-0.547,-0.0924)} \\
 14176 & {CB[A]} & 1.73 & -0.186 & {(-0.254,-0.119)} &
{(-0.273,-0.0913)} \\
 14176 & {AB[C]} & 2.09 & -0.226 & {(-0.382,-0.0955)} &
{(-0.426,-0.0649)} \\
 14176 & {CD[B]} & 2.13 & 0.292 & {(0.109,0.573)} &
{(0.0753,0.661)} \\
 14176 & {AD[B]} & 2.13 & 0.409 & {(0.200,0.597)} &
{(0.126,0.649)} \\
 2237 & {AD[B]} & 1.01 & -0.0801 & {(-0.155,-0.0318)} &
{(-0.174,-0.0220)} \\
 2237 & {BD[A]} & 1.18 & -0.128 & {(-0.239,-0.0527)} &
{(-0.264,-0.0369)} \\
 2237 & {AC[D]} & 1.37 & 0.297 & {(0.108,0.586)} &
{(0.0729,0.682)} \\
 2237 & {BC[D]} & 1.4 & 0.316 & {(0.122,0.607)} &
{(0.0820,0.704)}
\enddata
\tablecomments{The first column gives the abbreviated lens name (the
full names appear in the first column of Table \ref{tab:lens-info}).
The next two columns list the image pair label and the separation
between the images in arcseconds. The letter in brackets in column two indicates the third image needed to compute time delay ratios.  The last three columns present
data computed from our numerical simulations, using the galaxy
sample of \citet{Bender-ellip}. The fourth column gives the median
value of the differential time delay in units of $\tau_0 \theta_E^2$ (cf. Table 2), and the fifth and sixth columns give the 95\% and 99\% confidence intervals of this same quantity.}
\end{deluxetable*}


\begin{deluxetable*}{lllrcc}
\tablewidth{0pt} \tabletypesize{\small} \tablecaption{Median values
and confidence intervals for time delay ratios \label{tab:tdel-rat}}
\tablehead{
Lens & Image & $d_1$    & Median                    & 95\% Conf. & 99\% Conf. \\
Name & Pair  & (arcsec) & $\Delta t_1 / \Delta t_2$ & Interval   &
Interval }
\startdata
0128 & AB & 0.14 & 0.107 & (0.0631, 0.230) & (0.0499, 0.334) \\
0128 & AD & 0.27 & 9.31 & (4.34, 15.9) & (2.99, 20.0) \\
0128 & CD & 0.42 & 3.82 & (2.48, 7.73) & (2.13, 10.4) \\
0128 & CB & 0.5 & 41.7 & (27.1, 101.) & (25.2, 160.) \\
0230 & AB & 0.74 & 0.0477 & (0.0330, 0.0942) & (0.0310, 0.124) \\
0230 & CD & 1.46 & 0.818 & (0.667, 1.17) & (0.633, 1.29) \\
0230 & AD & 1.64 & 22.9 & (11.3, 47.5) & (8.58, 70.0) \\
0230 & CB & 1.65 & 13.9 & (7.41, 33.7) & (6.46, 46.0) \\
0414 & A1A2 & 0.41 & 0.0175 & (0.00996, 0.0288) & (0.00794, 0.0318) \\
0414 & BA2 & 1.71 & 57.2 & (34.6, 100.) & (31.4, 126.) \\
0414 & A1C & 1.96 & 261. & (92.7, 997.) & (70.2, 1650.) \\
0414 & BC & 2.13 & 3.08 & (1.57, 7.38) & (1.25, 9.64) \\
0712 & AB & 0.17 & 0.00961 & (0.00448, 0.0170) & (0.00359, 0.0193) \\
0712 & CB & 0.91 & 104. & (58.4, 220.) & (50.9, 269.) \\
0712 & CD & 1.18 & 4.12 & (1.83, 10.7) & (1.43, 14.0) \\
0712 & AD & 1.25 & 996. & (310., 4170.) & (215., 6910.) \\
0810 & AB & 0.18 & 0.0230 & (0.0132, 0.0360) & (0.0111, 0.0408) \\
0810 & CB & 0.69 & 43.5 & (27.7, 75.7) & (24.4, 89.9) \\
0810 & AD & 0.84 & 205. & (76.9, 761.) & (57.6, 1370) \\
0810 & CD & 0.85 & 3.38 & (1.59, 8.43) & (1.27, 11.1) \\
0924 & AD & 0.69 & 0.210 & (0.128, 0.422) & (0.105, 0.564) \\
0924 & AC & 1.18 & 4.77 & (2.37, 7.84) & (1.77, 9.52) \\
0924 & BD & 1.46 & 6.22 & (4.12, 9.90) & (3.56, 12.0) \\
0924 & BC & 1.53 & 2.50 & (1.61, 4.41) & (1.40, 5.45) \\
1004 & BA & 3.73 & 0.0256 & (0.0123, 0.0533) & (0.00918, 0.0683) \\
1004 & BD & 11.44 & 46.8 & (20.8, 121.) & (16.3, 213.) \\
1004 & CA & 11.84 & 29.4 & (18.3, 51.5) & (15.7, 63.0) \\
1004 & CD & 14.38 & 2.03 & (1.38, 3.30) & (1.23, 4.03) \\
1115 & A1A2 & 0.48 & 0.0179 & (0.00849, 0.0397) & (0.00650, 0.0514) \\
1115 & A1B & 1.67 & 60.2 & (26.4, 141.) & (20.1, 232.) \\
1115 & CB & 1.99 & 1.83 & (1.28, 3.08) & (1.12, 3.83) \\
1115 & CA2 & 2.16 & 108. & (51.1, 323.) & (44.2, 494.) \\
1330 & AB & 0.43 & 0.0267 & (0.0155, 0.0426) & (0.0135, 0.0474) \\
1330 & CB & 1.53 & 37.5 & (23.5, 64.4) & (21.0, 74.2) \\
1330 & CD & 1.64 & 2.29 & (0.937, 4.89) & (0.766, 6.72) \\
1330 & AD & 1.65 & 102. & (41.8, 323.) & (30.5, 591.) \\
1555 & AB & 0.09 & 0.0202 & (0.0116, 0.0327) & (0.00928, 0.0361) \\
1555 & CB & 0.35 & 49.4 & (30.6, 86.2) & (27.6, 107.) \\
1555 & AD & 0.4 & 168. & (66.3, 553.) & (51.7, 1030.) \\
1555 & CD & 0.42 & 3.21 & (1.52, 7.86) & (1.21, 10.2) \\
1608 & AC & 0.87 & 0.229 & (0.175, 0.320) & (0.162, 0.351) \\
1608 & BC & 1.51 & 6.48 & (3.69, 14.3) & (3.20, 19.0) \\
1608 & AD & 1.69 & 13.1 & (7.29, 20.8) & (6.72, 23.2) \\
1608 & BD & 2.00 & 2.21 & --- & --- \\
1933 & 4\_3 & 0.46 & 0.340 & (0.207, 0.638) & (0.166, 0.829) \\
1933 & 4\_6 & 0.63 & 2.94 & (1.57, 4.83) & (1.20, 5.99) \\
1933 & 1\_3 & 0.9 & 8.93 & (5.41, 17.4) & (4.67, 22.6) \\
1933 & 1\_6 & 0.91 & 3.67 & (2.21, 7.20) & (1.89, 9.26) \\
2026 & A1A2 & 0.33 & 0.0640 & (0.0350, 0.156) & (0.0280, 0.230) \\
2026 & A1C & 0.83 & 15.6 & (6.41, 28.5) & (4.27, 35.4) \\
2026 & BC & 1.19 & 3.64 & (2.22, 6.94) & (1.89, 8.77) \\
2026 & BA2 & 1.28 & 64.0 & (34.4, 143.) & (29.8, 208.) \\
2033 & A1A2 & 0.72 & 0.0856 & (0.0484, 0.187) & (0.0384, 0.247) \\
2033 & A1C & 1.54 & 11.7 & (5.34, 20.7) & (4.04, 26.0) \\
2033 & BA2 & 2.01 & 23.9 & (13.9, 47.8) & (12.2, 65.7) \\
2033 & BC & 2.13 & 2.86 & (1.84, 5.49) & (1.60, 6.88) \\
0911 & BA & 0.48 & 0.495 & (0.262, 0.956) & (0.171, 1.23) \\
0911 & BC & 0.62 & 2.01 & (1.03, 3.70) & (0.754, 5.73) \\
0911 & DC & 2.96 & 118. & (73.6, 292.) & (55.1, 551.) \\
0911 & DA & 3.08 & 292. & (156., 741.) & (126., 912.) \\
1131 & BA & 1.19 & 0.856 & (0.589, 1.25) & (0.548, 1.42) \\
1131 & CA & 1.26 & 1.17 & (0.800, 1.70) & (0.696, 1.82) \\
1131 & BD & 3.14 & 37.5 & (15.2, 120.) & (11.4, 199.) \\
1131 & CD & 3.18 & 35.2 & (14.1, 112.) & (10.5, 170.) \\
1251 & BA & 0.44 & 0.278 & (0.148, 0.558) & (0.114, 0.824) \\
1251 & BC & 0.7 & 3.59 & (1.79, 6.73) & (1.18, 8.75) \\
1251 & DA & 1.72 & 64.9 & (34.8, 147.) & (30.4, 211.) \\
1251 & DC & 1.77 & 20.7 & (11.0, 50.3) & (9.43, 68.9) \\
1422 & AB & 0.5 & 0.261 & (0.177, 0.376) & (0.157, 0.412) \\
1422 & CB & 0.82 & 3.83 & (2.65, 5.64) & (2.42, 6.31) \\
1422 & AD & 1.25 & 25.3 & (11.1, 62.4) & (8.60, 106.) \\
1422 & CD & 1.29 & 7.02 & (3.13, 18.1) & (2.48, 29.7) \\
2045 & AB & 0.28 & 0.155 & (0.0809, 0.293) & (0.0735, 0.323) \\
2045 & CB & 0.56 & 6.34 & (3.29, 12.2) & (3.02, 12.8) \\
2045 & AD & 1.91 & 508. & (185., 1890.) & (130., 3580.) \\
2045 & CD & 1.93 & 75.6 & (30.4, 234.) & (23.1, 406.) \\
0435 & CB & 1.53 & 0.899 & (0.604, 1.34) & (0.502, 1.62) \\
0435 & AB & 1.59 & 1.11 & (0.744, 1.66) & (0.618, 2.00) \\
0435 & CD & 1.85 & 2.30 & (1.23, 4.08) & (0.997, 5.02) \\
0435 & AD & 1.88 & 2.19 & (1.18, 3.99) & (0.982, 5.03) \\
12531 & BC & 0.77 & 0.967 & (0.670, 1.42) & (0.587, 1.69) \\
12531 & AC & 0.78 & 1.09 & (0.754, 1.63) & (0.638, 1.99) \\
12531 & BD & 0.91 & 2.42 & (1.28, 5.12) & (1.04, 7.47) \\
12531 & AD & 1.02 & 3.83 & (1.80, 10.0) & (1.44, 13.4) \\
14113 & CD & 1.13 & 0.515 & (0.285, 0.968) & (0.221, 1.23) \\
14113 & CB & 1.38 & 2.79 & (1.41, 6.07) & (1.15, 8.95) \\
14113 & AD & 1.41 & 2.19 & (1.45, 3.63) & (1.25, 4.41) \\
14113 & AB & 1.42 & 1.54 & (0.887, 2.27) & (0.777, 2.64) \\
1413 & AB & 0.76 & 0.702 & (0.389, 1.18) & (0.314, 1.50) \\
1413 & AC & 0.87 & 1.77 & (1.00, 3.08) & (0.832, 3.77) \\
1413 & DC & 0.91 & 1.41 & (0.896, 2.07) & (0.790, 2.47) \\
1413 & DB & 0.96 & 2.21 & (1.46, 3.67) & (1.26, 4.47) \\
14176 & CB & 1.73 & 0.625 & (0.431, 0.959) & (0.366, 1.17) \\
14176 & AB & 2.09 & 1.86 & (1.23, 2.95) & (1.04, 3.56) \\
14176 & CD & 2.13 & 2.56 & (1.34, 4.82) & (1.08, 6.27) \\
14176 & AD & 2.13 & 1.37 & (0.693, 2.75) & (0.564, 3.65) \\
2237 & AD & 1.01 & 0.627 & (0.417, 0.922) & (0.350, 1.07) \\
2237 & BD & 1.18 & 1.60 & (1.09, 2.41) & (0.935, 2.88) \\
2237 & AC & 1.37 & 3.55 & (1.80, 6.69) & (1.45, 8.74) \\
2237 & BC & 1.4 & 2.31 & (1.25, 4.72) & (1.02, 6.70) \\
\enddata
\tablecomments{
The first three columns are the same as those in Table \ref{tab:tdel-scaled}. The last three columns present data computed from our numerical simulations, using the galaxy sample of \citet{Bender-ellip}. The fourth column gives the median value of the time delay ratio $\Delta t_1 / \Delta t_2$. The subscripts on $\Delta t$ refer to the time delay of the labeled image pair ($\Delta t_1)$ and that for the closest neighboring pair ($\Delta t_2$). The fifth and sixth columns give the 95\% and 99\% confidence intervals of the time delay ratio.
}
\end{deluxetable*}

\end{document}